\documentclass[twocolumn,showpacs,floatfix,
reprint,
superscriptaddress,
amsmath,amssymb,
aps,
pre]{revtex4-2}

\usepackage{graphicx}
\usepackage{dcolumn}
\usepackage{amsmath}
\usepackage{bm}
\usepackage{appendix}
\usepackage{multirow}
\usepackage[hidelinks]{hyperref}
\usepackage[mathlines]{lineno}

\usepackage{inconsolata} 
\usepackage[usenames,dvipsnames]{color} 
\usepackage{listings}

\newcommand\underrel[2]{\mathrel{\mathop{#2}\limits_{#1}}}
\begin{document}

\preprint{APS/123-QED}

\title{Critical dynamics of long-range quantum disordered systems}

\author{Weitao Chen}
\affiliation{Department of Physics, National University of Singapore, Singapore.}
\affiliation{MajuLab, CNRS-UCA-SU-NUS-NTU International Joint Research Unit, Singapore.}
\affiliation{Centre for Quantum Technologies, National University of Singapore, Singapore.}
\author{Gabriel Lemari\'e}%
\email{lemarie@irsamc.ups-tlse.fr}
\affiliation{MajuLab, CNRS-UCA-SU-NUS-NTU International Joint Research Unit, Singapore.}
\affiliation{Centre for Quantum Technologies, National University of Singapore, Singapore.}
\affiliation{Laboratoire de Physique Théorique, Université de Toulouse, CNRS, UPS, France.}
\author{Jiangbin Gong}%
\affiliation{Department of Physics, National University of Singapore, Singapore.}
\affiliation{MajuLab, CNRS-UCA-SU-NUS-NTU International Joint Research Unit, Singapore.}
\affiliation{Centre for Quantum Technologies, National University of Singapore, Singapore.}


\date{\today}

\begin{abstract}
Long-range hoppings in quantum disordered systems are known to yield quantum multifractality, whose features can go beyond the characteristic properties associated with an Anderson transition. Indeed, critical dynamics of long-range quantum systems can exhibit anomalous dynamical behaviours distinct from those at the Anderson transition in finite dimensions. In this paper, we propose a phenomenological model of wave packet expansion in long-range hopping systems. We consider both their multifractal properties and the algebraic fat tails induced by the long-range hoppings. Using this model, we analytically derive the dynamics of moments and Inverse Participation Ratios of the time-evolving wave packets, in connection with the multifractal dimension of the system. To validate our predictions, we perform numerical simulations of a Floquet model that is analogous to the power law random banded matrix ensemble. Unlike the Anderson transition in finite dimensions, the dynamics of such systems cannot be adequately described by a single parameter scaling law that solely depends on time. Instead, it becomes crucial to establish scaling laws involving both the finite-size and the time.  Explicit scaling laws for the observables under consideration are presented.  Our findings are of considerable interest towards applications in the fields of many-body localization and Anderson localization on random graphs, where long-range effects arise due to the inherent topology of the Hilbert space.
\end{abstract}

\maketitle


\section{Introduction}
The study of eigenstate transitions in quantum-disordered systems has attracted a strong interest recently \cite{Sachdev_1999}. One celebrated example is the Anderson transition arising from the interplay between interference effects and disorder, which separates a phase where quantum states are localized from a phase where states are delocalized \cite{PhysRev.109.1492, RevModPhys.80.1355, abrahams201050}. At the Anderson transition, a property called multifractality emerges as a consequence of strong and scale invariant spatial fluctuations of the states, intermediate between localization and delocalization \cite{mandelbrot1974intermittent,mandelbrot1982fractal,falconer2004fractal}. Given the importance of the Anderson transition, multifractal properties have been extensively investigated, both theoretically and experimentally, in finite-dimension \cite{PhysRevLett.102.106406,PhysRevLett.103.155703,PhysRevB.84.134209, PhysRevA.80.043626, PhysRevLett.105.090601,PhysRevLett.101.255702} and in random matrix ensembles \cite{PhysRevE.54.3221,PhysRevB.62.7920,PhysRevLett.97.046803}. Recently, it was discovered that quantum multifractality can be observed not only at critical points but also in phases called extended non-ergodic \cite{PhysRevLett.113.046806, Kravtsov_2015}. For example, the many-body localized phase has been shown to have multifractal properties on the Hilbert space \cite{PhysRevLett.123.180601, PhysRevB.102.014208,PhysRevB.104.024202}. The emergence of such non-ergodic extended phases have also been described in random matrix ensembles \cite{Kravtsov_2015, 10.21468/SciPostPhys.6.1.014, PhysRevResearch.2.043346, von2019non, Truong_2016, PhysRevE.98.032139, Monthus_2017, Amini_2017, PhysRevB.103.104205, kravtsov2020localization, PhysRevResearch.2.043346, 10.21468/SciPostPhys.11.2.045}, on the Cayley tree \cite{PhysRevB.94.184203, biroli2018delocalization, kravtsov2018non, parisi2019anderson}, in Floquet systems \cite{PhysRevE.81.066212,PhysRevE.97.010101, PhysRevB.103.184309, 10.21468/SciPostPhys.4.5.025, PhysRevB.93.104504, 10.21468/SciPostPhys.12.3.082} or in the presence of long-range correlations of disorder \cite{PhysRevB.106.L020201, roy2020localization}. 

Quantum multifractality can be characterized by the moments $P_q$ of order $q$ of eigenstate amplitudes:
\begin{equation}\label{eq:Pq}
    \langle P_q\rangle =\langle \sum_i|\Psi_\alpha(i)|^{2q}\rangle\sim N^{-D_q(q-1)},
\end{equation}
where the sum is over the $N$ sites (indexed by $i$) of the system, with the eigenstate amplitudes $|\Psi_\alpha(i)|^{2}$ normalized as $\sum_i|\Psi_\alpha(i)|^{2}=1$. $\langle \rangle$ denotes an averaging over disorder and eigenstates in a certain energy window. An algebraic scaling of $\langle P_q\rangle$ with $N$ defines a multifractal dimension $D_q$.
$D_q=1$ indicates an ergodic delocalized behavior, while $D_q=0$ is a signature of localization. These behaviors are generally observed at a sufficiently large scale, e.g. $N \gg \Lambda$ the correlation or localization volume.
Remarkably, $0<D_q<1$ indicates scale-invariant multifractal behaviors, whose full characterization is based on a spectrum of multifractal dimensions \cite{RevModPhys.80.1355}. Multifractal eigenstates thus occupy an extensive region which is however an algebraically vanishing fraction of the system: this is why they are called ``non-ergodic delocalized'' \cite{PhysRevLett.113.046806, Kravtsov_2015}, in contrast to egodic delocalized states which occupy a finite fraction of the system.   

As a characteristic property of the Anderson transition in finite dimensions \cite{castellani1986multifractal, PhysRevB.62.7920, PhysRevLett.103.155703, PhysRevLett.105.046403, PhysRevB.84.134209}, quantum multifractality is even richer in the presence of long-range hoppings.
One well-known example is the Power-law Random Banded Matrix (PRBM) model \cite{PhysRevE.54.3221, PhysRevB.62.7920}. Similarly, Floquet models, particularly the Ruijsenaars-Schneider ensemble \cite{PhysRevE.84.036212,PhysRevE.85.046208}, exhibit intriguing properties of quantum multifractality. There, long-range hoppings introduce anomalous properties beyond typical features at the Anderson transition in finite dimension. For example, they result in an unusually large critical regime \cite{PhysRevLett.87.056601}, can break a fundamental symmetry of the multifractal spectrum \cite{PhysRevResearch.3.L022023}, and induce correlation-induced localization \cite{PhysRevB.99.104203}. In this article, we explore how long-range hoppings also give rise to anomalous \textit{dynamical} properties.

In terms of detecting the Anderson transition,  solely observing the expansion of a wave packet already serves as a convenient tool. Indeed, this has been extensively studied both theoretically and experimentally \cite{doi:10.1143/JPSJ.66.314, PhysRevA.80.043626, PhysRevA.94.033615, PhysRevLett.105.090601, PhysRevLett.101.255702, PhysRevLett.124.186601}. Precisely at the Anderson transition point, the wavepacket expansion exhibits an anomalous diffusion behavior that lies between localization and diffusion \cite{doi:10.1143/JPSJ.66.314, PhysRevA.80.043626, PhysRevLett.105.090601, PhysRevLett.101.255702, PhysRevLett.124.186601}, which needs more careful quantitative analysis of the wave packet spatial profile to see the impact of multifractality.  Other observables such as the return probability or the coherent back and forward scattering peaks are more useful to study the multifractal properties of the eigenstates and the Cantor eigen-spectrum \cite{PhysRevLett.69.695, PhysRevB.82.161102, Kravtsov_2011, hopjan2022scale, altshuler2023random, PhysRevA.95.041602, martinez2022coherent, PhysRevA.100.043612}. 

Investigation of the quantum dynamics in long-range hopping systems \cite{PhysRevLett.79.1959,PhysRevE.86.056215, PhysRevE.86.021136, Kravtsov_2011, PhysRevB.82.161102, PhysRevLett.69.695, PhysRevB.82.161102, altshuler2023random} in connection with multifractality is more challenging, insofar as the dynamics is strongly affected by the algebraic tails induced by the long-range hoppings (analogous to l\'evy flights \cite{PhysRevE.86.021136}), as we will show in this paper (see also \cite{PhysRevB.71.235112, Amini_2017}).  In particular, the strong boundary effects caused by the algebraic tails present a severe challenge in computational studies.  As shown in this work, it is not possible to circumvent these strong boundary effects by increasing the system size. 
In other words, one cannot reach a regime where expansion of a wave packet
is not affected by finite-size effects, thus requiring more scaling analysis than in the case of the Anderson transition studied both theoretically and experimentally in different platforms \cite{PhysRevLett.101.255702, PhysRevLett.105.090601, PhysRevA.100.043612, PhysRevA.94.033615, doi:10.1143/JPSJ.66.314}. Interestingly, in the cases we consider, it is necessary to take into account systematically the boundary effects via a scaling in time, in addition to the system size. Indeed, the focus of this paper is on understanding the subtle critical dynamical behaviors induced by long-range hopping via a two-parameter (time and size) scaling approach. Some results were already discussed in Refs. \cite{PhysRevB.71.235112, Amini_2017} using different approaches. This paper distinguishes itself from these studies by providing a coherent description of long-range coupling effects based on an unified model of wave packet propagation in these systems.

Studies of the dynamics of quantum systems are generally computationally expensive. In comparison, simulating the dynamical counterpart of Floquet kicked systems can be made more efficient, as is clear in the kicked rotor dynamics implemented via Fast Fourier Transforms \cite{santhanam2022quantum}. Besides their computational efficiency, Floquet kicked systems also exhibit rich dynamical behaviors such as dynamical localization \cite{CHIRIKOV1979263,IZRAILEV1990299}, or Floquet time crystals \cite{PhysRevLett.117.090402}.
In this work, we employ a Floquet kicked model with algebraically long-range hoppings and eigenstates with multifractal properties \cite{PhysRevLett.94.244102} to simulate numerically the critical dynamics in such long-range hopping systems. We propose scaling laws whose scaling parameters include both time and system size, for different observables, based on a general and simple phenomenological model of wave packet expansion in the type of systems considered. Our analytical and numerical results demonstrate distinct dynamical behaviors depending on the observables considered.

As an outlook, we note that algebraic fat tails in time evolving wave packets are also relevant to studies of quantum dynamics on various graphs of infinite effective dimension, such as Anderson localization in random graphs \cite{PhysRevResearch.2.012020, PhysRevResearch.3.L022023} or the Hilbert space of a many-body localized system \cite{PhysRevB.97.214205}. The Hilbert space of these systems have network structure where the number $N_r$ of sites at distance $r$ from the localization center of a wave packet grows exponentially, therefore the exponential decay of the wave packet with distance $r$ can be regarded as an algebraic behavior as a function of $N_r$.  Under the new coordinate $N_r$, important localization measures like the inverse participation ratio can be more easily studied since the network structure is simplified to 1-D. Hence, our findings hold potential relevance in this context, which has recently gathered significant attention.

 The rest of the paper is organized as follows. In Sec.~\ref{sec2}, we introduce the kicked Floquet model we consider and discuss its multifractal properties. In Sec.~\ref{sec:R0}, we recall the temporal behavior and finite-size dependence of the return probability $\langle R_0\rangle$ and generalize these known results to higher moments $\langle {R_0}^q(t)\rangle$. In Sec.~\ref{sec3}, we propose a general phenomenological model of wave packet expansion in long-range hopping systems with multifractal properties, based on both analytical arguments and numerical observations. In Sec.~\ref{sec4}, we derive from the phenomenological model the dynamics and time and size scaling laws for two other important types of observables (some may be accessible experimentally): the average $k$-th moments of a wave packet $\langle p^k\rangle$ and the $q$-th Inverse Participation Ratio $\langle P_q (t) \rangle$. We present numerical results that validate our predictions. We conclude our study in Sec. \ref{sec5}.

\section {The multifractal kicked rotor model}\label{sec2}
 In this article, we investigate a variant of the quantum kicked rotor \cite{CHIRIKOV1979263,IZRAILEV1990299} that we call the multifractal kicked rotor (MKR) model, with Hamiltonian \cite{PhysRevLett.94.244102}
\begin{equation}\label{eq2}
     \mathcal{H}=\frac{p^{2}}{2}+KV(q)\sum_{n}\delta(t-n),
\end{equation}
where
\begin{equation}\label{eq3}
\begin{split}
\begin{aligned}
V(q) =     \begin{cases}
       &\ln (q/\pi),q\in [0,\pi), \\
       &\ln (2-q/\pi),q\in [\pi,2\pi) ,
    \end{cases} 
    \end{aligned}
\end{split}
\end{equation}
 and $V(q+2\pi)=V(q)$. Hamiltonian Eq. (\ref{eq2}) yields a Floquet operator $U=\exp(-p^2/2\hbar)\exp(-iKV(q)/\hbar)$, which can be quantized in a truncated Hilbert space with dimension $N$ with $p=P\hbar$, $P$ an integer between $-\frac{N}{2}$ and $\frac{N}{2}-1$, and $q=\frac{2\pi Q-\varepsilon}{N}$, $Q$ an integer between $1$ and $N$ satisfying periodic boundary conditions in both $P$ and $Q$. However, note that we have assigned the value of $\varepsilon=1$ for $q\in [\pi,2\pi)$ (i.e., when $Q=1,\dots, \frac{N}{2}$), while for other values of $q$, we have set $\varepsilon=0$ to prevent numerical divergence. Such slight shifts break the symmetry of the kicking potential Eq. (\ref{eq3}) with respect to the axis $q=\pi$, consequently, the time-reversal symmetry of the Hamiltonian is broken. The phases corresponding to the kinetic energy $\Phi_{P}\equiv P^2\hbar/2$ are pseudo-random phases when $\hbar$ is irrational with $2 \pi$ \cite{PhysRevLett.49.509, birkhoff1931proof,PhysRevA.29.1639}. Here, we consider $\Phi_{P}$ 
 as fully-random phases, uniformly distributed over $[0,2\pi)$. Without loss of generality, we set $\hbar=1$ in the rest of the paper. We can therefore treat $p$ and $P$  as the same variable, and we will no longer use the notation $P$ in the following.
 
 The Floquet operator can be explicitly expressed in the momentum space using a discrete Fourier transform as 
\begin{equation}\label{eq:Uop}
\begin{split}
      U_{pp^{\prime}}=e^{-i \Phi_p}\sum_{Q=1}^{N}F_{pQ}e^{-iKV(2\pi Q/N)}F_{Qp^{\prime}}^{-1},
\end{split}
\end{equation}
where $F_{pQ}=\frac{1}{\sqrt{N}}e^{2i\pi pQ/N}$. Due to the singular behavior of $V(q)$ when $q\rightarrow 0 \; (2\pi)$, the amplitudes of the matrix elements of $U_{pp^{\prime}}$ decay as 
\begin{equation}\label{eq5}
    |U_{pp^{\prime}}|\sim\frac{1}{|p-p^{\prime}|}
\end{equation}
 for large $|p-p^{\prime}|$ (note that there is another higher-order singularity at $q=\pi$ which can be negelected), see \cite{PhysRevLett.94.244102} and App.~\ref{appendixA} for more details). 
In  App.~\ref{appendixB}, we characterize the multifractal properties of the MKR model, and in particular extract the multifractal dimension $D_2=0.71$ for $K=10$ by analyzing the system size dependence of eigenstate moments numerically. Another Floquet system, the Ruijsenaars-Schneider model with similar long-range hopping amplitudes, has been extensively studied for its multifractal properties \cite{PhysRevE.85.046208}, spectral statistics \cite{giraud2004intermediate}, and rich dynamics \cite{PhysRevE.86.056215}. 
 
\section{Return probability $R_0$}\label{sec:R0}

\begin{figure}
\includegraphics[width=0.5\textwidth]{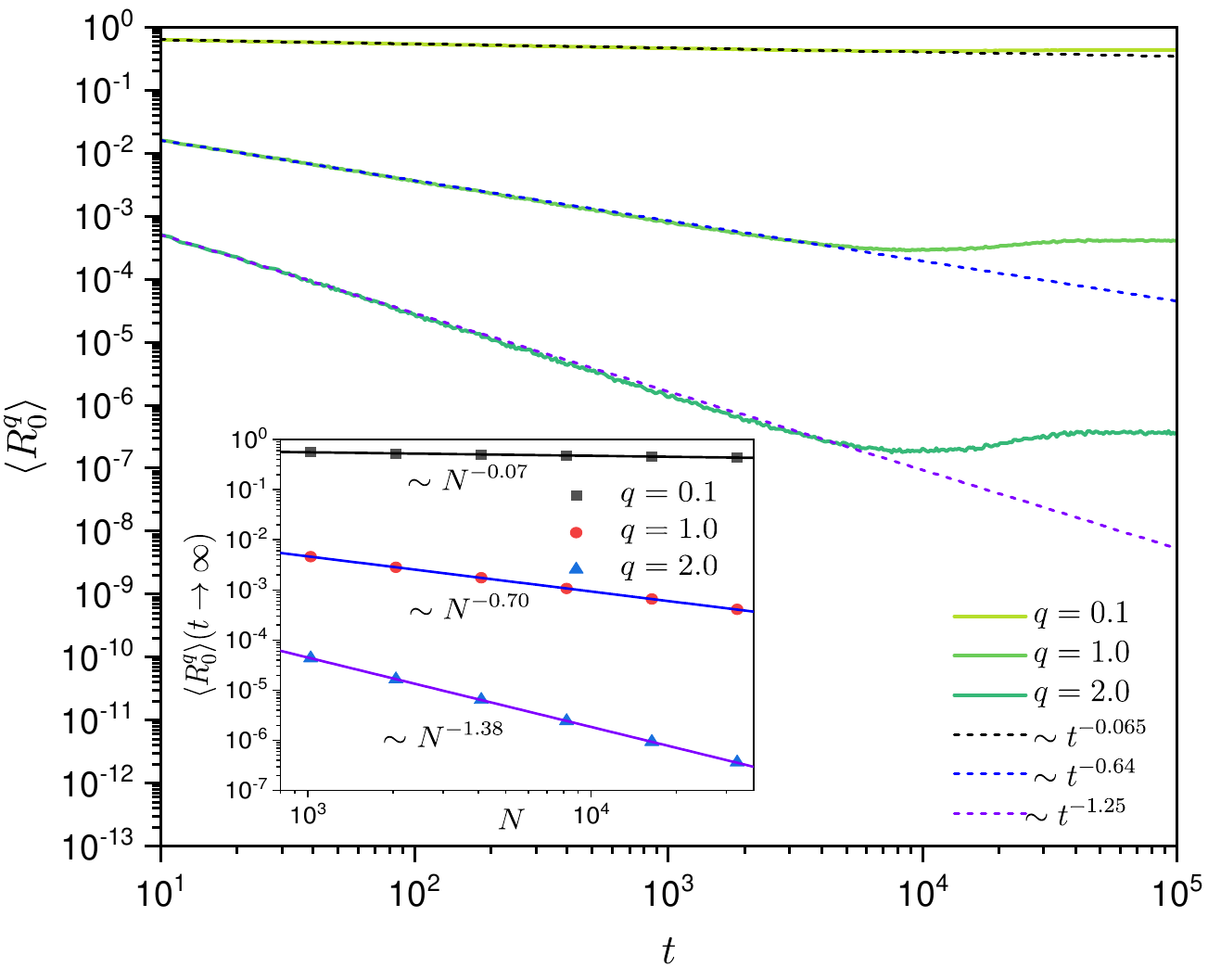}
\caption{\label{figR0_t(R0inf_N)} Dynamics of  $\langle R_0^q \rangle$ in the MKR model Eq.~\eqref{eq2} for $K=10$ and $q=0.1$, $1.0$ and $2.0$ from top to bottom. The dashed lines indicate fits with an algebraic law, validating Eq.~\eqref{eq6} for $t<t^{*}_N$ with $D_2^{\mu}=0.64$. The system size is $N=2^{15}$. Inset: Finite size scaling of saturation values of $\langle R_0^q(t\rightarrow\infty)\rangle$, verifying Eq.~\eqref{eq7} for $t>t^{*}_N$ with $D_2^{\psi}=0.70$. The multifractal dimensions $D_2^{\mu} \approx D_2^{\psi} \approx D_2=0.71$, as expected for this type of system \cite{altshuler2023random}. Numerical results have been averaged over $4800$ random phase configurations.}
\end{figure}

\begin{figure}
\includegraphics[width=0.5\textwidth]{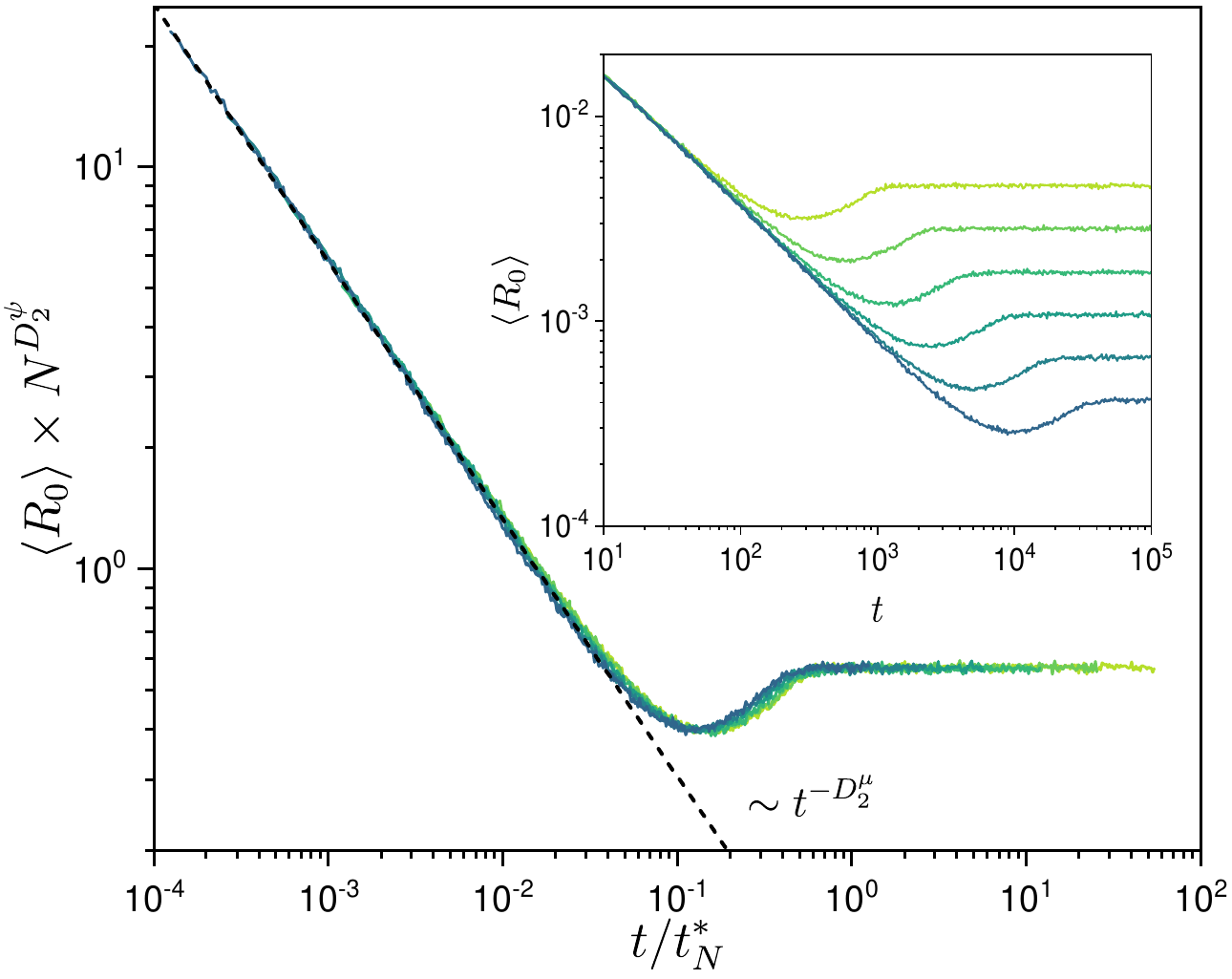}
\caption{\label{figScaled_R0_t}  Demonstration of the scaling property Eq.~\eqref{eq:scaR0} for the return probability $R_0$ in the MKR model Eq.~\eqref{eq2}: the data corresponding to different system sizes $N=2^{10},2^{11},...,2^{15}$ and different times $t \in [10, 10^5]$ all collapse onto a single scaling curve when $\langle R_0\rangle \times N^{D_2^{\psi}}$ is plotted as a function of the scaled time $t/t^*_N$. The dashed line indicates a fit by the power-law $ \langle R_0\rangle\sim t^{-D_2^{\mu}}$ with $D_2^\mu \approx 0.64$. Inset: corresponding raw data $\langle R_0\rangle $. Curves from top to bottom correspond to system sizes $N=2^{10},2^{11},...,2^{15}$, respectively. Results have been averaged over $4800$ disorder configurations with the kicked strength $K=10$.}
\end{figure}

Before describing the behavior of observables which are significantly affected by the long-range hoppings introduced above, we recall the properties of a dynamical observable, the return probability $R_0$, which has been extensively investigated as a characteristic signature of quantum multifractality \cite{PhysRevLett.61.593, CHALKER1990253, Kravtsov_2011, PhysRevB.82.161102, PhysRevA.100.043612}. Starting from an initial condition $\psi(p,t=0) = \delta_{p,0}$, $R_0$ is defined as $R_0 \equiv \vert \psi(p=0,t)\vert^2$.
As a result of the Cantor eigenspectrum, $R_0$ decays as a power law in the time domain, $\langle R_0\rangle \sim t^{-D_2^{\mu}}$ where $D_2^{\mu}$ is the multifractal dimension of the spectral measure \cite{PhysRevLett.69.695, PhysRevB.76.235119,PhysRevB.82.161102, Kravtsov_2011, altshuler2023random}. 

In our study, higher moments $\langle R_0^q\rangle$ with $q>0$ will play a key role. Due to narrow distributions of large wave function amplitudes $|\psi|^2$ in such systems, see \cite{PhysRevResearch.3.L022023} and Appendix \ref{appendixD}, the power-law decay of $\langle R_0\rangle$ can be simply generalized to $\langle R_0^q\rangle$ with $q>0$ as, 
\begin{equation}\label{eq6}
\begin{split}
         \langle R_0^q\rangle \sim t^{-qD_2^{\mu}},
 \end{split}
 \end{equation}
as illustrated in Fig.~\ref{figR0_t(R0inf_N)}.
 
 On the other hand, in a finite system of size $N$, there exists a characteristic time scale $t^{*}_N$ after which $R_0$ reaches a finite stationary value, equal to the inverse participation ratio $\langle P_2\rangle$, Eq.~\eqref{eq:Pq}, i.e. $ \langle R_0(t\rightarrow\infty)\rangle =\langle P_2\rangle \sim N^{-D_2^{\psi}}$,
where $D_2^{\psi}$ is the spatial multifractal dimension of the eigenstates \cite{PhysRevB.71.235112, PhysRevLett.79.1959}. Similarly, we find that the size dependence of $\langle R_0^q\rangle$ at large times follows:
\begin{equation}\label{eq7}
    \langle R_0^q(t\rightarrow\infty)\rangle \sim N^{-qD_2^{\psi}}\;.
\end{equation}
Therefore, the characteristic time $t^*_N$ should scale as 
\begin{equation}\label{eq:tstar}
    t^*_N\sim N^{D_2^{\psi}/D_2^{\mu}} \; .
\end{equation}
$t^*_N$ reduces to the Heisenberg time (inverse of the mean level spacing $2 \pi/N$) for systems with $D_2^{\psi}=D_2^{\mu}$ \cite{hopjan2022scale,PhysRevB.82.161102, Kravtsov_2011}. Combining the above relations, we can infer the following two parameter scaling behavior for $R_0$, namely,
\begin{equation}
\begin{split}
    \langle R_0^{q}(t,N)\rangle =N^{-qD_2^{\psi}}g(t/t^*_N).
\end{split} \label{eq:scaR0}
\end{equation}

The numerical data for the MKR verify the above scaling relations. Results presented in Fig.~\ref{figR0_t(R0inf_N)} validate Eq.~(\ref{eq6}) and Eq.~(\ref{eq7}). By fitting the corresponding data, we extract the multifractal dimensions $D_2^{\mu}=0.64$ and  $D_2^{\psi}=0.70$. In Fig.~\ref{figScaled_R0_t}, the collapse of $R_0$ onto a single scaling curve when $\langle R_0\rangle N^{D_2^{\psi}}$ is plotted as a function of $ t/t^*_N$ confirms the validity of the proposed scaling law Eq.~\eqref{eq:scaR0}. In App.~\ref{appendixC}, we show similar scaling properties for $\langle R_0^q\rangle$ with $q=0.1$ and $2$. Similar scaling properties for $\langle R_0 \rangle$ (i.e. $q=1$) have been observed in Ref.~\cite{hopjan2022scale} in both single-particle and many-body quantum systems.

\section{Phenomenological model for the expansion of a wave packet in multifractal systems with algebraic long-range hoppings}\label{sec3}

\begin{figure}
\includegraphics[width=0.5\textwidth]{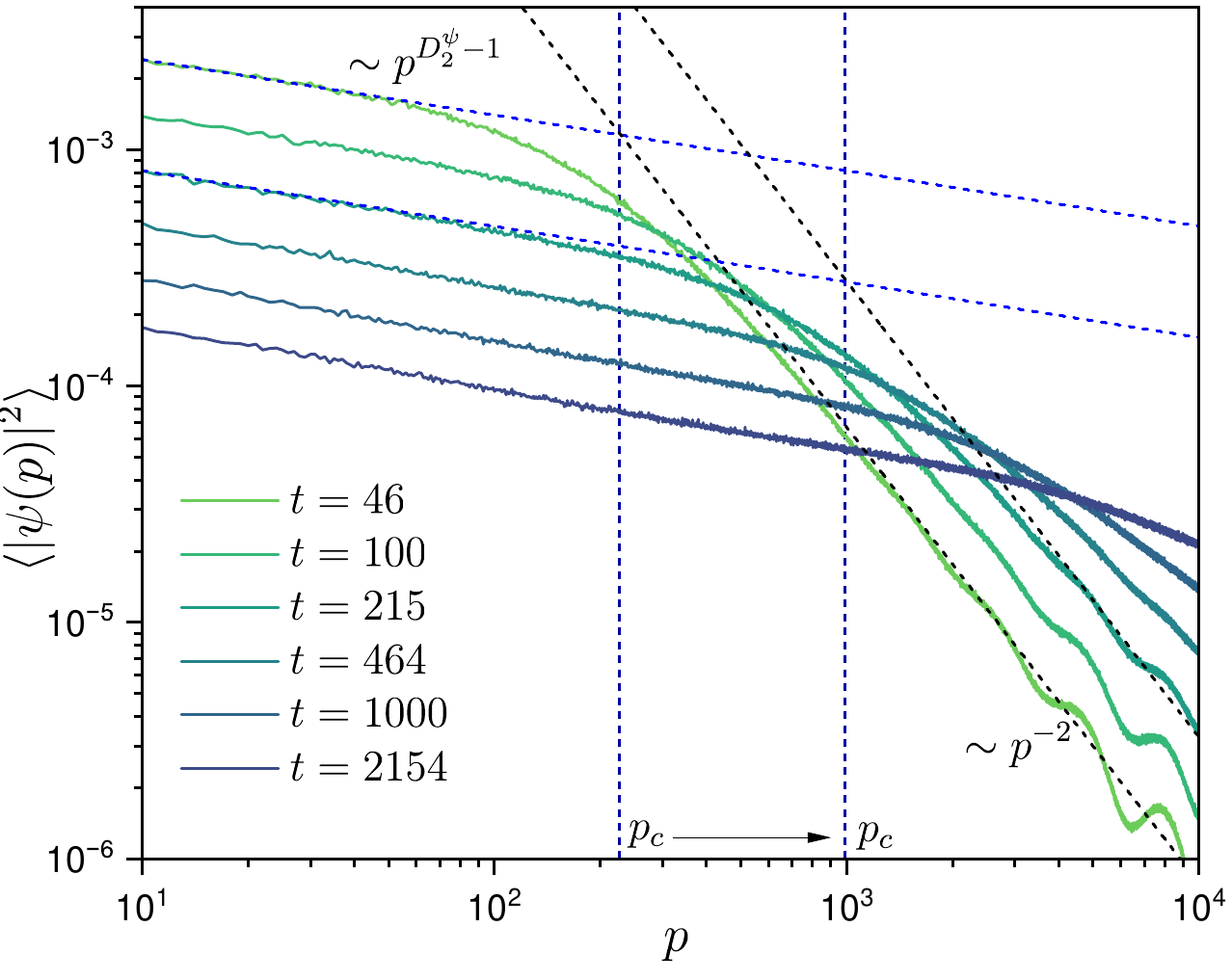}
\caption{\label{figPsi2_p} Average probability distribution of wave packets $\langle|\psi(p,t)|^2\rangle$ at different times for the MKR model with initial condition $\psi(p,t=0) = \delta_{p,0}$. The dashed lines show the two power-law behaviors corresponding to Eq.(\ref{eq9}) and the multifractal wave-front $p_c$. Results have been averaged over $4800$ disorder configurations with the kicked strength $K=10$ for system size $N=2^{15}$.}
\end{figure}

We shall now describe the rich and subtle effects of algebraic long-range hoppings on the critical dynamics of a wave packet, effects that cannot be characterized using the widely used return probability.
We construct in this section a phenomenological model, based on known analytical results and simple arguments such as wave packet normalization, and validate this model by numerical simulations using the MKR model Eq.~\eqref{eq2}. Here we restrict our analysis to the regime $p>0$, since we expect similar scaling behavior for $p<0$, as the wave packet is initialized as $\psi(p,t=0) = \delta_{p,0}$.

Starting from a wave packet initialized at a single site $p=0$, long-range hoppings will induce a power-law tail of the wave packet. This tail is primarily determined by the hopping elements before any interference effects induced by multifractality occur. If the long-range hoppings follow the behavior described in Eq. (\ref{eq5}), then the tail of the wave packet behaves as $\langle|\psi(p)|^2\rangle\sim p^{-2}$, so called the l\'evy flight tail \cite{PhysRevE.86.021136}. However, in the vicinity of the site $p=0$ where the wave packet was initialized, a non-trivial power-law decay $\langle|\psi(p)|^2\rangle\sim p^{D_2^{\psi}-1}$ dynamically emerges, which is controlled by the spatial correlation dimension $D_2^{\psi}$ of the wavefunction \cite{PhysRevLett.79.1959, CHALKER1990253, PhysRevA.100.043612, PhysRevE.86.021136}.

Fig.~\ref{figPsi2_p} represents the averaged probability distribution of a wave packet initialized at $p=0$, $\langle|\psi(p,t)|^2\rangle$ at different times, for the MKR model Eq.~\eqref{eq2}. Two distinct power-law decays with $p$ are clearly visible: a fast decay $\langle|\psi(p,t)|^2\rangle\sim p^{-2}$ at large $p\gg p_c$, and a slower decay $\langle|\psi(p,t)|^2\rangle\sim p^{D_2^{\psi}-1}$ close to the initial condition $p\ll p_c$. The crossover scale $p_c$ has a non-trivial dependence on time which we will describe in the following. It is equivalent to the characteristic scale mentioned in \cite{PhysRevLett.61.593, CHALKER1990253}, which distinguishes the scaling behaviors of the density correlation function in the position-frequency representation, specifically between the large and small position regimes. 
Crucially for our study, we also observe that other moments of wave packet amplitudes, $\langle|\psi(p,t)|^{2q}\rangle$ with $q>0$, obey a similar behavior, see Eq.~\eqref{eq9} below. The distributions for different $q$ thus share the same shape, in particular the same $p_c$ (see App.~\ref{appendixD} for more details). 

Based on these observations, we propose the following phenomenological model for the average probability distributions of the generalized wave packets $\langle|\psi(p,t)|^{2q}\rangle$ for $q>0$,
\begin{equation}\label{eq9}
\begin{split}
\begin{aligned}
\langle|\psi(p,t)|^{2q}\rangle =     \begin{cases}
       & \langle R_0^q\rangle p^{-q\mu}, \quad 1\leq p< p_c, \\
       &B\left[\frac{p}{p_c}\right]^{-q\lambda}, \quad p_c< p\leq \frac{N}{2} ,
    \end{cases} 
    \end{aligned}
\end{split}
\end{equation}
where $\lambda$ is the exponent of the power-law tail at large $p\gg p_c$ ($\lambda=2$ in the MKR model), $\mu=1-D_2^{\psi}$ is the exponent of the power law decay at small $p\ll p_c$, related to the multifractal dimension $D_2^{\psi}$, and $B=\langle R_0^q\rangle[p_c]^{-q\mu}$. Note that our model is valid only above a microscopic cutoff taken as $p_\text{min}=1$ here. This cutoff usually corresponds to the mean free path, see e.g.~\cite{PhysRevA.100.043612}. In the following, we will neglect contributions below this cutoff, which are not of our interest here.

The crossover scale $p_c$ between the two power-law regimes can be interpreted as a multifractal wave-front. Its dynamics and finite-size scaling play an important role in the following. They can be understood simply by invoking normalization of the wave packet $||\psi||^{2} \equiv \sum_{p=-\frac{N}{2}}^{p=\frac{N}{2}-1} \vert \psi(p,t)\vert^2\simeq 2\int_1^{\frac{N}{2}}\vert \psi(p,t)\vert^2 \, dp$, where we have taken into account the fact that the wave packet is symmetric with respect to the origin and neglected contributions below the cutoff $p< 1$. Therefore:
\begin{equation}
\begin{split}
\begin{aligned}
   1=||\psi||^{2}\simeq&2\left[\int_{1}^{p_{c}}\langle R_0\rangle p^{-\mu}dp+\int_{p_c}^{\frac{N}{2}}B\left(\frac{p}{p_c}\right)^{-\lambda }dp\right]\\
    =&\frac{2\langle R_0\rangle}{1-\mu}\left({p_c}^{1-\mu}-1\right)\\
    +&\frac{2\langle R_0\rangle{p_c}^{-\mu}}{1-\lambda}\left[{p_c}^{\lambda}\left(\frac{N}{2}\right)^{1-\lambda}-p_c\right].
\end{aligned}
\end{split}
\end{equation}
The previous expression can be simplified, using $\mu=1-D_2^{\psi}$, as
\begin{equation}
\begin{split}
\begin{aligned}
    \left(\frac{1}{D_2^{\psi}}+\frac{1}{\lambda-1}\right){p_c}&^{D_2^{\psi}}-\frac{1}{\lambda-1}{p_c}^{\lambda+D_2^{\psi}-1}\left(\frac{N}{2}\right)^{1-\lambda}\\
    &\simeq\frac{1}{2\langle R_0\rangle}+\frac{1}{D_2^{\psi}}.
\end{aligned}
\end{split}
\end{equation}
The second term in the left-hand side of the above equality vanishes when $N\rightarrow\infty$ if $\lambda>1$, as is the case in the MKR model considered. Using $\langle R_0\rangle\sim t^{-D_2^{\mu}}$ for $t<t^*_N$, Eq.~\eqref{eq:tstar}, we get the following dynamical behavior of $p_c$:
 \begin{equation}\label{eq12}
     p_c\sim t^{\frac{D_2^{\mu}}{D_2^{\psi}}},\quad\quad(t\ll t^{*}_N)
 \end{equation}
In the limit of large times $t\gg t^*_N$, substituting $R_0(t\rightarrow\infty)=P_2\sim N^{-D_2^{\psi}}$, one gets
 \begin{equation}\label{eq13}
     p_c(t\rightarrow\infty)\sim N.\quad\quad(t\gg t^{*}_N)
 \end{equation}
As said above, we have shown in  App.~\ref{appendixD} that the multifractal wave-front $p_c$ is the same for generalized wave packets $\langle|\psi(p,t)|^{2q}\rangle$ with different $q$.

\section{Two parameter scaling in size and time for critical quantum dynamics of algebraically long-range systems}\label{sec4}

\begin{table*}
\begin{tabular}{|c|c|c|c|c|c|}
\hline
\multirow{2}{*}{Observable} & \multicolumn{2}{c|}{$\langle p^k\rangle$} & %
    \multicolumn{3}{c|}{$\langle P_q\rangle$} \\
\cline{2-6}

 & $k<\lambda-1$ & $k>\lambda-1$ & $0<q<\frac{1}{\lambda}$ & $\frac{1}{\lambda}<q<\frac{1}{1-D_2^{\psi}}$ &$q>\frac{1}{1-D_2^{\psi}}$\\
\hline
 finite-time dependence&$t^{k(D_2^{\mu}/D_2^{\psi})}$ & $t^{(\lambda-1)D_2^{\mu}/D_2^{\psi}}$&$ t^{q(\lambda-1)D_2^{\mu}/D_2^{\psi}}$ & $t^{(1-q)D_2^{\mu/}D_2^{\psi}}$ &$t^{-qD_2^{\mu}}$\\
\hline
 finite-size dependence&$N^k$ &$N^k$ &$N^{1-q}$ & $N^{1-q}$ &$N^{-qD_2^{\psi}}$\\
\hline
 scaling law&\multicolumn{2}{c|}{$N^kg(t/t^*_N)$} &\multicolumn{2}{c|}{$N^{1-q}g(t/t^*_N)$}   &$N^{-qD_2^{\psi}}g(t/t^*_N)$\\
\hline
\end{tabular}
\caption{\label{tab1}Summary of the analytical predictions of the finite-time and finite-size dependence of the dynamics of the $k$-th moments $\langle p^k\rangle$ and $q$-th inverse participation ratios $\langle P_q(t)\rangle$.}
\end{table*}

In this section, we employ the phenomenological model introduced in Eq.~(\ref{eq9}) to derive the critical dynamics dependent on time and size, in terms of the average $k$-th moments of a wave packet $\langle p^k\rangle$. The observation of $\langle p^k\rangle$ is possible in cold atom systems \cite{PhysRevA.80.043626, PhysRevLett.101.255702} and ultrasound experiments \cite{PhysRevLett.103.155703}, thus making it an experimentally accessible observable. Furthermore, we examine the $q$-th Inverse Participation Ratio $\langle P_q (t)\rangle$, which is a significant quantity in standard multifractal analysis; see Ref.~\cite{PhysRevE.86.056215} for more information. Based on these dynamical observables, we propose scaling laws that are dependent on both time and size. In Tab.~(\ref{tab1}), we summarize the analytical predictions of the finite-time and size-dependent dynamics for both observables. The MKR model of Eq.~(\ref{eq2}) is used to numerically verify the predicted critical dynamics and their respective scaling laws.

\subsection{$k$-th moment $\langle p^k\rangle$ of a wave packet}
\begin{figure}
\includegraphics[width=0.5\textwidth]{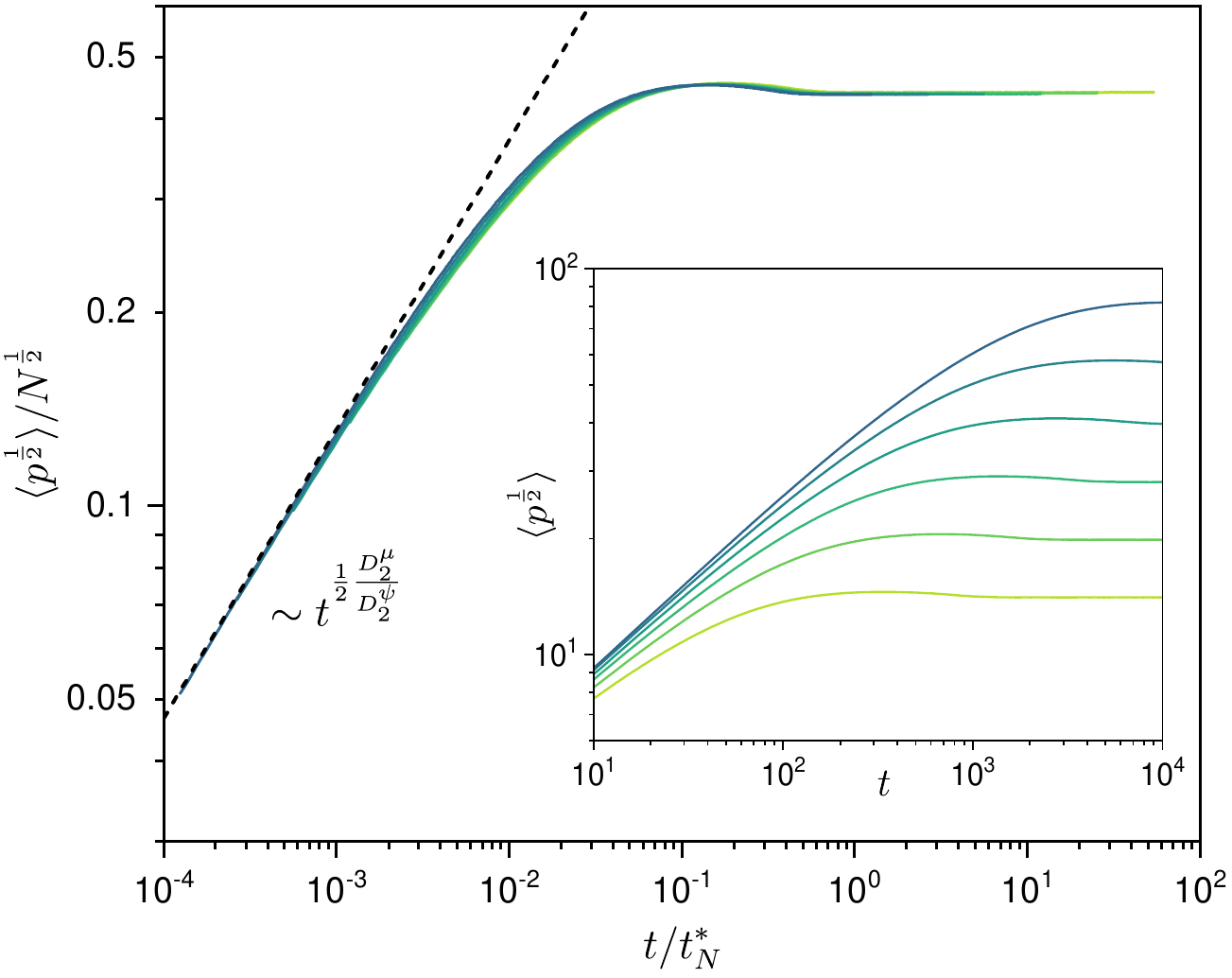}
    \caption{\label{figScaled_x05_t} Demonstration of the scaling property Eq.~\eqref{eq17} for the moment of the wave packet  $\langle p^{\frac{1}{2}}\rangle$ in the MKR model Eq.~\eqref{eq2}: the data corresponding to different system sizes $N=2^{10},2^{11},...,2^{15}$ and different times $t \in [10, 10^5]$ all collapse onto a single scaling curve when $\langle p^{\frac{1}{2}}\rangle/N^{\frac{1}{2}}$ is plotted as a function of the scaled time $t/t^*_N$. The dashed line indicates a fit by the power-law $ \langle p^{\frac{1}{2}}\rangle\sim t^{\frac{1}{2}\frac{D_2^{\mu}}{D_2^{\psi}}}$. Inset: corresponding raw data $\langle  p^{\frac{1}{2}}\rangle $. Curves from top to bottom correspond to system sizes $N=2^{10},2^{11},...,2^{15}$, respectively. Results have been averaged over $4800$ disorder configurations with the kicked strength $K=10$.}
\end{figure}

\begin{figure}
\includegraphics[width=0.5\textwidth]{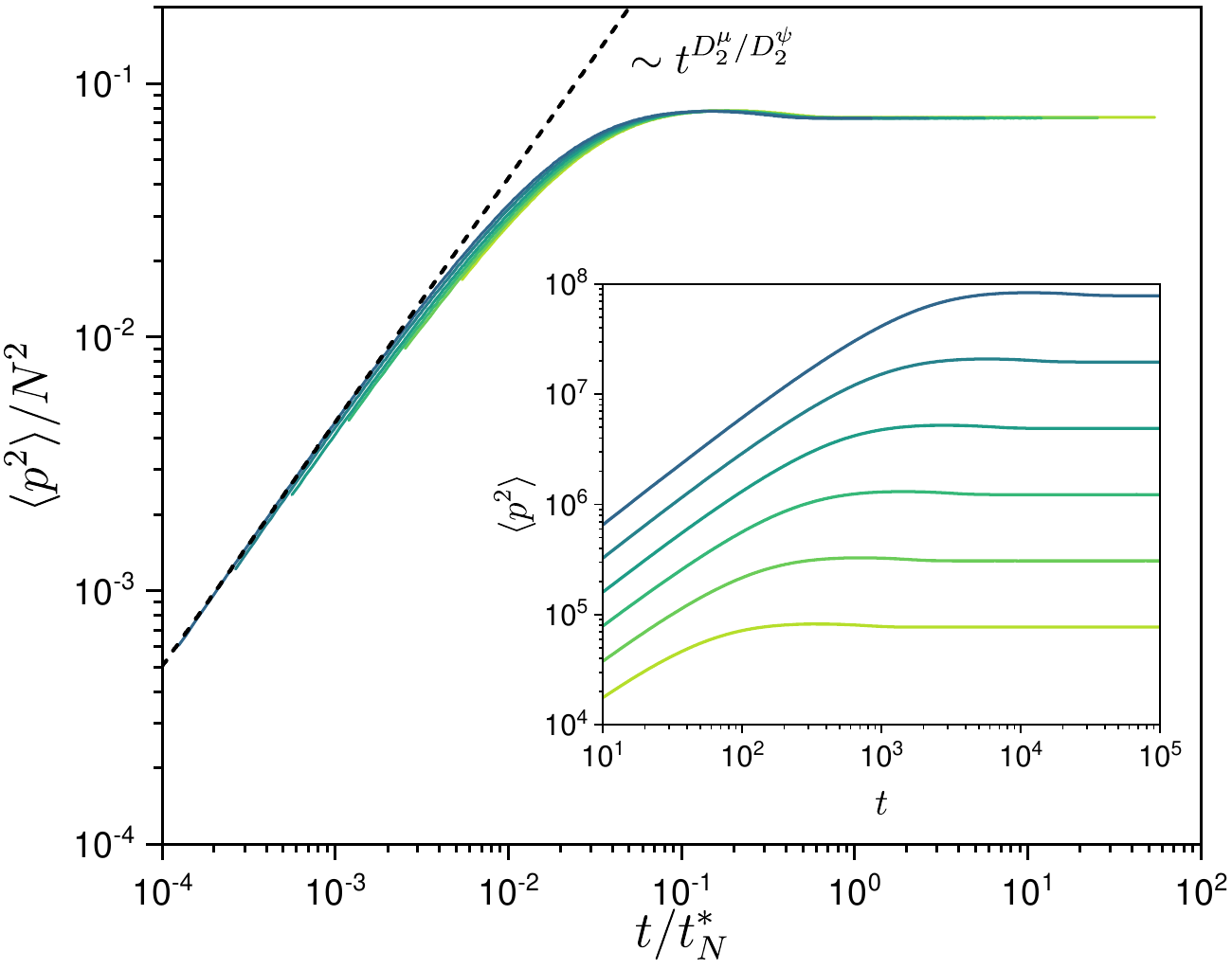}
    \caption{\label{figScaled_x2_t}  Demonstration of the scaling property Eq.~\eqref{eq17} for the moment of the wave packet  $\langle p^{2}\rangle$ in the MKR model Eq.~\eqref{eq2}: the data corresponding to different system sizes $N=2^{10},2^{11},...,2^{15}$ and different times $t \in [10, 10^5]$ all collapse onto a single scaling curve when $\langle p^{2}\rangle/N^{2}$ is plotted as a function of the scaled time $t/t^*_N$. The dashed line indicates a fit by the power-law $ \langle p^{2}\rangle\sim t^{D_2^{\mu}/D_2^{\psi}}$. Inset: corresponding raw data $\langle  p^{2}\rangle $. Curves from top to bottom correspond to system sizes $N=2^{10},2^{11},...,2^{15}$, respectively. Results have been averaged over $4800$ disorder configurations with the kicked strength $K=10$.}
\end{figure}

The average $k$-th moments of a wave packet $\langle p^k\rangle$:
\begin{equation}
    \langle p^k\rangle=\sum_{p=-\frac{N}{2}}^{p=\frac{N}{2}-1}|p|^{k}\langle |\psi(p,t)|^2\rangle
\end{equation}
reflect the diffusive properties of a system. Note that we have defined the moments using an absolute value of $p$ since the wave packet is symmetric with respect to the origin $p=0$.
Based on the phenomenological model proposed in Eq.~(\ref{eq9}), 
\begin{equation} \label{eq:pk}\small 
\begin{split}
\begin{aligned}
    \langle p^{k}\rangle\approx &2\int_1^{\frac{N}{2}}\langle|\psi(p,t)|^2\rangle p^kdp\\ 
    =&\int_{1}^{p_{c}}2\langle R_0\rangle p^{k-\mu}dp+\int_{p_{c}}^{\frac{N}{2}}2Bp_{c}^{\lambda}p^{k-\lambda }dp\\
    =&\frac{2\langle R_0\rangle}{k+1-\mu}\left(p_{c}^{k+1-\mu}-1\right)\\
    +&\frac{2\langle R_0\rangle}{k+1-\lambda}\left[p_{c}^{\lambda-\mu}\left(\frac{N}{2}\right)^{k+1-\lambda}-p_{c}^{k+1-\mu}\right].
\end{aligned}
\end{split}
\end{equation}
Combining the time-dependent analysis of $p_c$ and $\langle R_0\rangle$ in Eq.~(\ref{eq:pk}), the time-dependent dynamics of $\langle p^k\rangle$ can be derived as
\begin{equation}\label{eq15}
\begin{aligned}
\langle p^{k}\rangle &\sim \langle R_0\rangle p_{c}^{k+1-\mu}+\langle R_0\rangle p_{c}^{\lambda-\mu}N^{k+1-\lambda}\\
   &\sim t^{k\frac{D_2^{\mu}}{D_2^{\psi}}}+t^{(\lambda-1)\frac{D_2^{\mu}}{D_2^{\psi}}}N^{k+1-\lambda}\; ,
\end{aligned}
\end{equation}
for $t\ll t^*_N$.

For $k<\lambda-1$, the second term of Eq.~(\ref{eq15}) vanishes when $N\rightarrow \infty$, yielding $\langle p^{k}\rangle\sim t^{k(D_2^{\mu}/D_2^{\psi})}$. This regime was previously investigated in the Fibonacci chain and Harper model in Ref.~\cite{PhysRevLett.79.1959}. Nevertheless, for $k>\lambda-1$, the second term dominates, contributing $\langle p^{k}\rangle\sim t^{(\lambda-1)D_2^{\mu}/D_2^{\psi}}$. For the MKR model, the power-law tail exponent $\lambda=2$, which yields $\langle p^{k}\rangle\sim t^{D_2^{\mu}/D_2^{\psi}}$ for $k>1$ and $\langle p^{k}\rangle\sim t^{kD_2^{\mu}/D_2^{\psi}}$ for $0<k<1$. The numerical results shown in Fig.~\ref{figScaled_x05_t} and Fig.~\ref{figScaled_x2_t} confirm such predictions. The diffusive exponents for $\langle p^{k}\rangle$ are independent of $k$ when $k>\lambda-1$, which is a non-trivial consequence of the power-law tail of the wave-packet induced by algebraic long-range hoppings. 

Furthermore, using the finite size scaling of $p_c \sim N$ and $R_0 \sim N^{-D_2^\psi}$ at large $t \gg t^*_N$, we can also derive the finite size scaling of $\langle p^k\rangle$:
\begin{equation}\label{eq16}
\begin{split}
   \langle p^k(t\rightarrow\infty)\rangle\sim N^k.
\end{split}
\end{equation}
Finally, a two parameter scaling law for $\langle p^k\rangle$ can be naturally proposed based on the time-dependence of Eq.~(\ref{eq15}) and the finite-size dependence of Eq.~(\ref{eq16}):
\begin{equation}\label{eq17}
\begin{split}
    \langle p^k(t,N)\rangle=N^kg(t/t^*_N).
\end{split}
\end{equation}
The presented numerical results in Fig.~\ref{figScaled_x05_t} and Fig.~\ref{figScaled_x2_t} demonstrate that the data for $\langle p^2\rangle$ of the MKR model adheres to the proposed scaling behavior. The data collapse onto a single scaling curve when $\langle p^2\rangle/N^2$ is plotted as a function of $t/t^*_N$. Additionally, in App.~\ref{appendixC}, we provide numerical data for $\langle p^3\rangle$ and $\langle p^5\rangle$, which confirms the validity of the aforementioned predictions.

\subsection{$q$-th inverse participation ratio $\langle P_q(t)\rangle$ of a wave packet}

\begin{figure}
\includegraphics[width=0.5\textwidth]{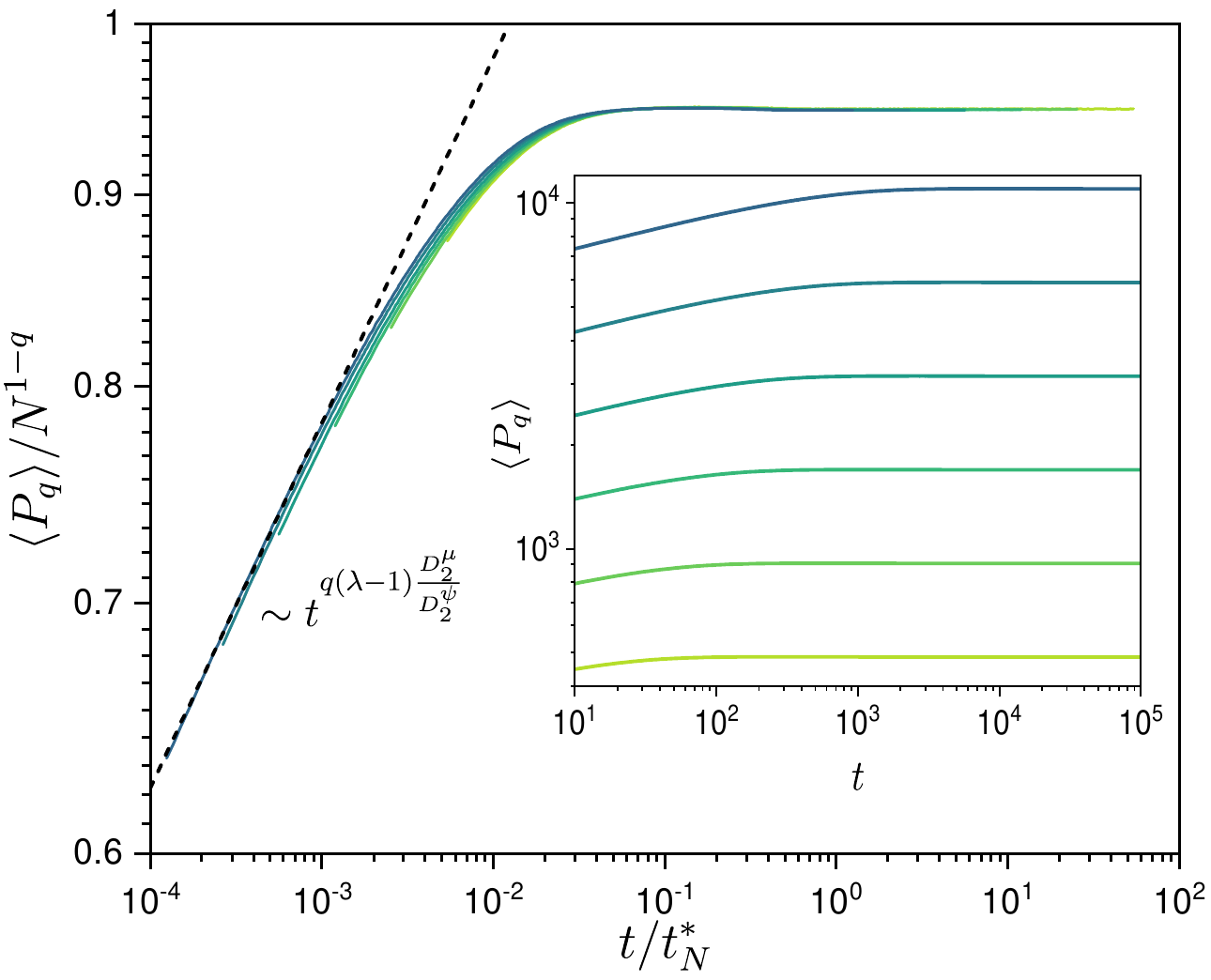}
\caption{\label{figScaled_P01_t} 
Demonstration of the scaling property Eq.~\eqref{eq24} for $\langle P_q\rangle$ in the MKR model Eq.~\eqref{eq2} when $q=0.1$: the data corresponding to different system sizes $N=2^{10},2^{11},...,2^{15}$ and different times $t \in [10, 10^5]$ all collapse onto a single scaling curve when $\langle P_q\rangle/N^{1-q}$ is plotted as a function of the scaled time $t/t^*_N$. The dashed line indicates a fit by the power-law $ \langle P_q\rangle \sim t^{q(\lambda-1)D_2^{\mu}/D_2^{\psi}}$. Inset: corresponding raw data $\langle  P_q\rangle $. Curves from top to bottom correspond to system sizes $N=2^{10},2^{11},...,2^{15}$, respectively. Results have been averaged over $4800$ disorder configurations with the kicked strength $K=10$.}
\end{figure}

\begin{figure}
\includegraphics[width=0.5\textwidth]{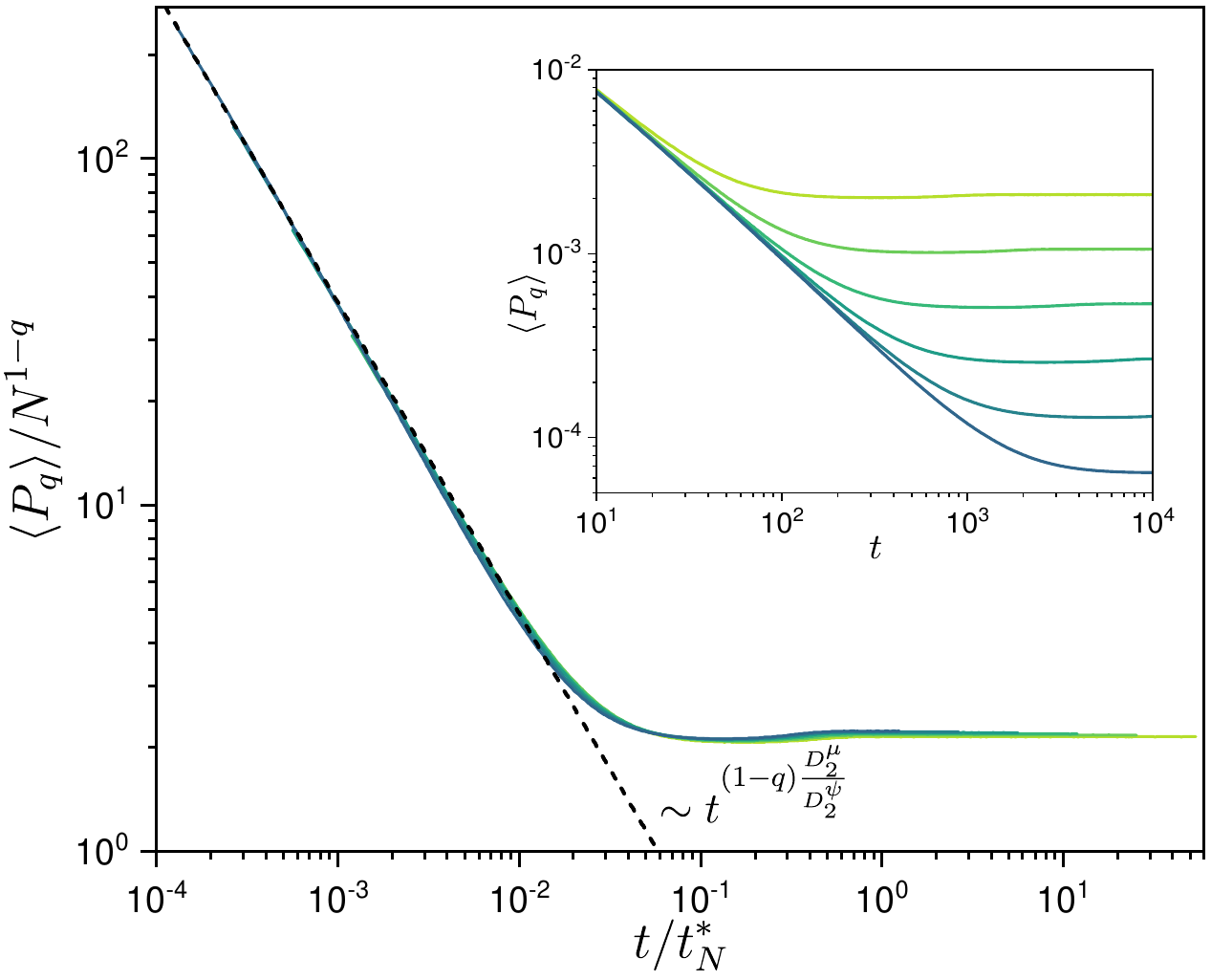}
\caption{\label{figScaled_P2_t} 
Demonstration of the scaling property Eq.~\eqref{eq24} for $\langle P_q\rangle$ in the MKR model Eq.~\eqref{eq2} when $q=2$: the data corresponding to different system sizes $N=2^{10},2^{11},...,2^{15}$ and different times $t \in [10, 10^5]$ all collapse onto a single scaling curve when $\langle P_q\rangle/N^{1-q}$ is plotted as a function of the scaled time $t/t^*_N$. The dashed line indicates a fit by the power-law $\langle P_q\rangle\sim t^{(1-q)D_2^{\mu}/D_2^{\psi}}$. Inset: corresponding raw data $\langle  P_q\rangle $. Curves from top to bottom correspond to system sizes $N=2^{10},2^{11},...,2^{15}$, respectively. Results have been averaged over $4800$ disorder configurations with the kicked strength $K=10$.}
\end{figure}

\begin{figure}
\includegraphics[width=0.5\textwidth]{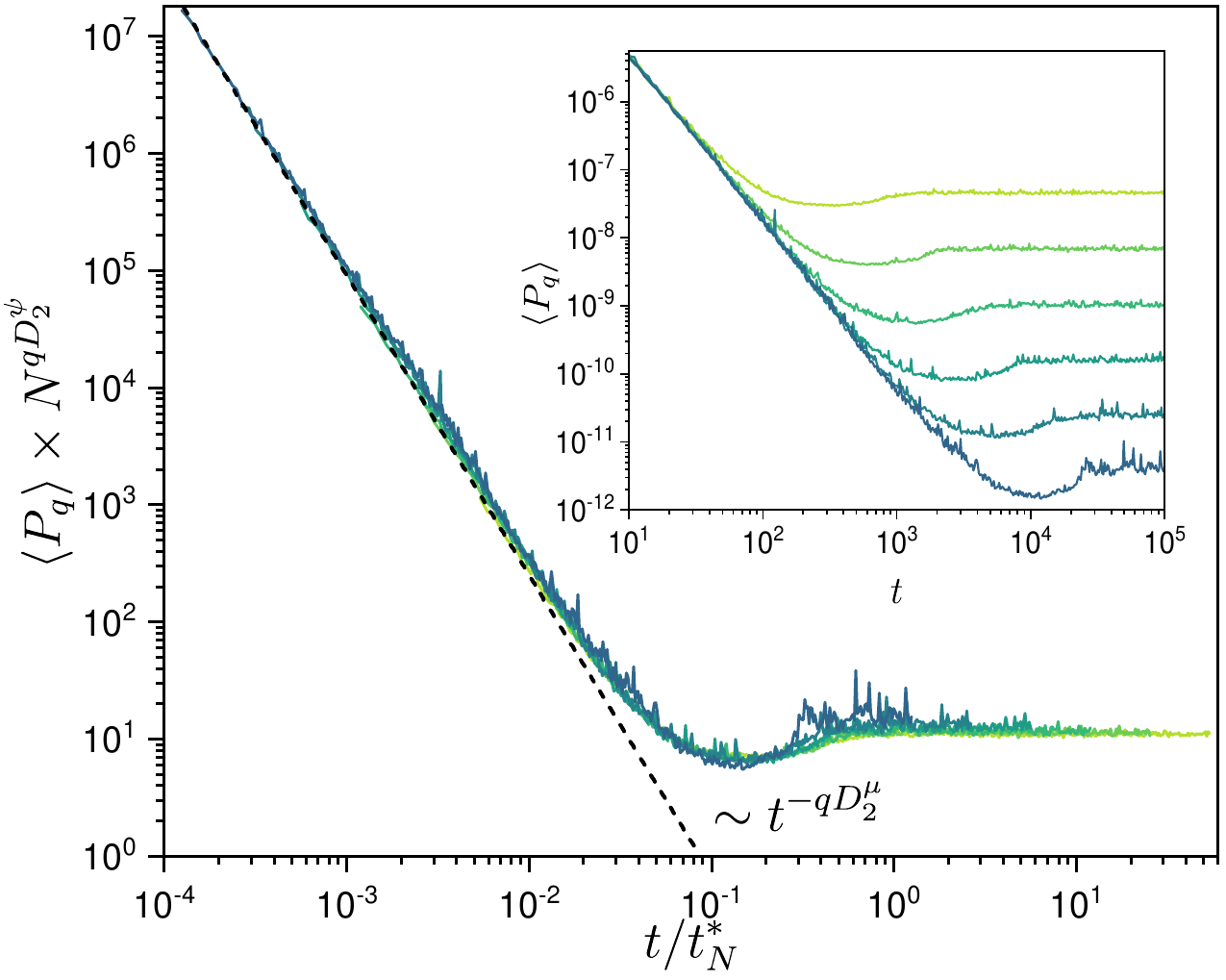}
\caption{\label{figScaled_P4_t}Demonstration of the scaling property Eq.~\eqref{eq24} for $\langle P_q\rangle$ in the MKR model Eq.~\eqref{eq2} when $q=4$: the data corresponding to different system sizes $N=2^{10},2^{11},...,2^{15}$ and different times $t \in [10, 10^5]$ all collapse onto a single scaling curve when $\langle P_q\rangle/N^{1-q}$ is plotted as a function of the scaled time $t/t^*_N$. The dashed line indicates a fit by the power-law $ \langle P_q\rangle \sim t^{-qD_2^{\mu}}$. Inset: corresponding raw data $\langle  P_q\rangle $. Curves from top to bottom correspond to system sizes $N=2^{10},2^{11},...,2^{15}$, respectively. Results have been averaged over $4800$ disorder configurations with the kicked strength $K=10$.}
\end{figure}

We now turn to another key observable for multifractal properties, the  generalized inverse participation ratios. As we are interested in the dynamics of a wave packet, we do not consider the $\langle P_q \rangle$ of the eigenstates, Eq.~\eqref{eq:Pq}, but the $\langle P_q (t)\rangle$ of the time-evolving wave packet at a certain instant $t$:
\begin{equation}
    \langle P_q (t)\rangle \equiv \left \langle \sum_{p=-\frac{N}{2}}^{p=\frac{N}{2}-1} |\psi(p,t)|^{2q} \right \rangle.
\end{equation}
We will study how $\langle P_q (t)\rangle$ scales with system size, but also characterize its temporal behavior. The scaling with system size of the moments $\langle P_q \rangle$ of eigenstates captures the multifractality of critical systems directly, exhibiting distinct algebraic behaviors for different values of $q$. By contrast, the moments $\langle P_q (t)\rangle$ for a time-evolving wave packet are different as they are non-equilibrium observables capturing the dynamical growth of the participation volume of the eigenstate (e.g., $\langle P_2(t) \rangle$ being the inverse volume occupied by the wave packet).    

Similar to the analysis of the average $k$-th moments  $\langle p^k\rangle$, $\langle P_q(t)\rangle$ can be calculated as 
\begin{equation}\small\label{eq18}
\begin{split}
\begin{aligned}
    \langle P_q(t)\rangle&=2\int_1^{\frac{N}{2}}\langle|\psi(p,t)|^{2q}\rangle dp\\ 
    &=\int_{1}^{p_{c}}2\langle R_0^{q}\rangle p^{-q\mu}dp+\int_{p_{c}}^{\frac{N}{2}}2B^{q}p_{c}^{q\lambda}p^{-q\lambda }dp\\
    &=\frac{2\langle R_0^{q}\rangle}{1-q\mu}\left(p_{c}^{1-q\mu}-1\right)\\
    &+\frac{2\langle R_0^{q}\rangle}{1-q\lambda}\left[p_{c}^{q(\lambda-\mu)}\left(\frac{N}{2}\right)^{1-q\lambda}-p_{c}^{1-q\mu}\right]\\
    &\sim t^{(1-q)\frac{D_2^{\mu}}{D_2^{\psi}}}+t^{q(\lambda-1)\frac{D_2^{\mu}}{D_2^{\psi}}}N^{1-q\lambda}.
\end{aligned}\end{split}
\end{equation}

 When $N\rightarrow \infty$, for $q>1/\lambda$, i.e., $q>\frac{1}{2}$ for $\lambda=2$, the second term of the right hand side of Eq.~(\ref{eq18}) vanishes, yielding the time-dependent decay $\langle P_q\rangle \sim t^{(1-q)D_2^{\mu}/D_2^{\psi}}$. Otherwise, for $q<1/\lambda$, i.e. $q<\frac{1}{2}$ for $\lambda=2$, the second term dominates, resulting in  a time-dependent increase $\langle P_q\rangle  \sim t^{q(\lambda-1)D_2^{\mu}/D_2^{\psi}}$. Applying a similar analysis for the finite-size saturation value at 
$t \gg t^*_N$ yields
 \begin{equation}
    \langle P_q(t\rightarrow\infty)\rangle\sim N^{-qD_2^{\psi}}N^{1-q\mu}\sim N^{1-q}.
\end{equation}

However, Eq.~(\ref{eq18}) is valid only if $p_{c}^{1-q\mu}-1>0$, i.e.,  $q<\frac{1}{1-D_2^{\psi}}$. If  $q>\frac{1}{1-D_2^{\psi}}$, the contribution of the multifractal wave-front in the integral is smaller than or close to $1$. The term $\frac{\langle R_0^{q}\rangle}{1-q\mu}(p_{c}^{1-q\mu}-1)$ is then dominated by $\langle R_0^{q}\rangle$, yielding
\begin{equation}\label{eq20}
\begin{split}
\begin{aligned}
    \langle P_q\rangle &\simeq \frac{2\langle R_0^{q}\rangle}{1-q\mu}+\frac{2\langle R_0^{q}\rangle}{1-q\lambda}p_{c}^{q(\lambda-\mu)}\left(\frac{N}{2}\right)^{1-q\lambda}\\
    &\sim t^{-qD_2^{\mu}}+t^{q(\lambda-1)\frac{D_2^{\mu}}{D_2^{\psi}}}N^{1-q\lambda},
\end{aligned}\end{split}
\end{equation}
and
\begin{equation}
    \langle P_q (t\rightarrow\infty)\rangle\sim  N^{-qD_2^{\psi}}.
\end{equation}

Eq.~(\ref{eq20}) shows another regime where $\langle P_q\rangle\sim t^{-qD_2^{\mu}}$ when $q>\frac{1}{1-D_2^{\psi}}$ and $q>1/\lambda$. Combining the derivations above, two scaling laws can be proposed as
\begin{equation}\label{eq24}
\begin{aligned}
    \begin{cases}
       \langle P_q(t,N)\rangle=N^{1-q}g(t/t^*_N),& 0<q<\frac{1}{1-D_2^{\psi}},\\
\langle P_q(t,N)\rangle=N^{-qD_2^{\psi}}g(t/t^*_N),& q>\frac{1}{1-D_2^{\psi}}.
    \end{cases} 
\end{aligned}
\end{equation}

Applying the above insights to the MKR model considered here, overall it is clear that there are three regimes where $P_q$ varies with distinct exponents. For $q<0.5$, $\langle P_q\rangle \sim t^{q(\lambda-1)D_2^{\mu}/D_2^{\psi}}$; for $0.5<q<\frac{1}{1-D_2^{\psi}}\approx3.3$, $\langle P_q\rangle\sim t^{(1-q)D_2^{\mu}/D_2^{\psi}}$ and for $q>\frac{1}{1-D_2^{\psi}}$, $\langle P_q\rangle \sim t^{-qD_2^{\mu}}$. Fig. (\ref{figScaled_P01_t}), (\ref{figScaled_P2_t}) and (\ref{figScaled_P4_t}) present the collapse of data for $q=0.1,2,4$ , corresponding to the three distinct dynamical regimes. The different dynamical exponents are in good agreement with the predictions and the collapse of the rescaled data confirms the validity of the proposed scaling laws, similar numerical observations are also reported in Ref.~\cite{PhysRevB.71.235112} with power-law banded Anderson model.

\section{Conclusion}\label{sec5}

In conclusion, we have presented a thorough investigation into the wave packet dynamics of disordered quantum critical systems, exploring the anomalous effects of long-range hoppings in the presence of multifractal properties eigenstates. 
Our study indicates that long-range hoppings can induce subtle and rich dynamical behaviors. For example, the wave packet variance may increase linearly with time, namely, $\langle p^2 \rangle \sim t$, which could be akin to a diffusive behavior \cite{Amini_2017} though the system is multifractal. The multifractal properties of the wave packet dynamics itself, as characterized by the $\langle P_q(t) \rangle$, are all related to the fractal dimension $D_2$, but,  according to the value of $q$, can have different scaling behaviors on system size $N$ and time $t$.

Algebraic tails of time-evolving wave packets appear generically in systems with long-range couplings, but also effectively in localization problems on graphs of infinite dimensionality, such as in many-body localization. For these systems, this work indicates that we cannot avoid finite size effects, and that these effects can be taken into account via a two-parameter scaling theory depending on time $t$ and the system size $N$.

It would be interesting to extend our study to either delocalized or localized phases of disordered systems with long-range hoppings, where the couplings decrease with a smaller or larger exponent than in the critical case, respectively (in one-dimension cases, the couplings would decrease with an exponent different from $-1$).

\begin{acknowledgments}

We wish to thank O. Giraud for
fruitful discussions. This study has been supported through
the EUR grant NanoX n° ANR-17-EURE-0009 in the framework of the "Programme des Investissements d’Avenir", research funding Grants No. ANR-17-CE30-0024, ANR-18-CE30-0017 and ANR-19-CE30-0013, and by the Singapore
Ministry of Education Academic Research Fund Tier I (WBS
No. R-144-000-437-114). We thank Calcul en Midi-Pyrénées
(CALMIP) and the National Supercomputing Centre (NSCC) of Singapore for computational resources and assistance.

\end{acknowledgments}

\appendix

\section{Power-law decay of the amplitudes of Floquet matrix elements $|U_{p,p^{\prime}}|$}\label{appendixA}

\begin{figure}
\includegraphics[width=0.4\textwidth]{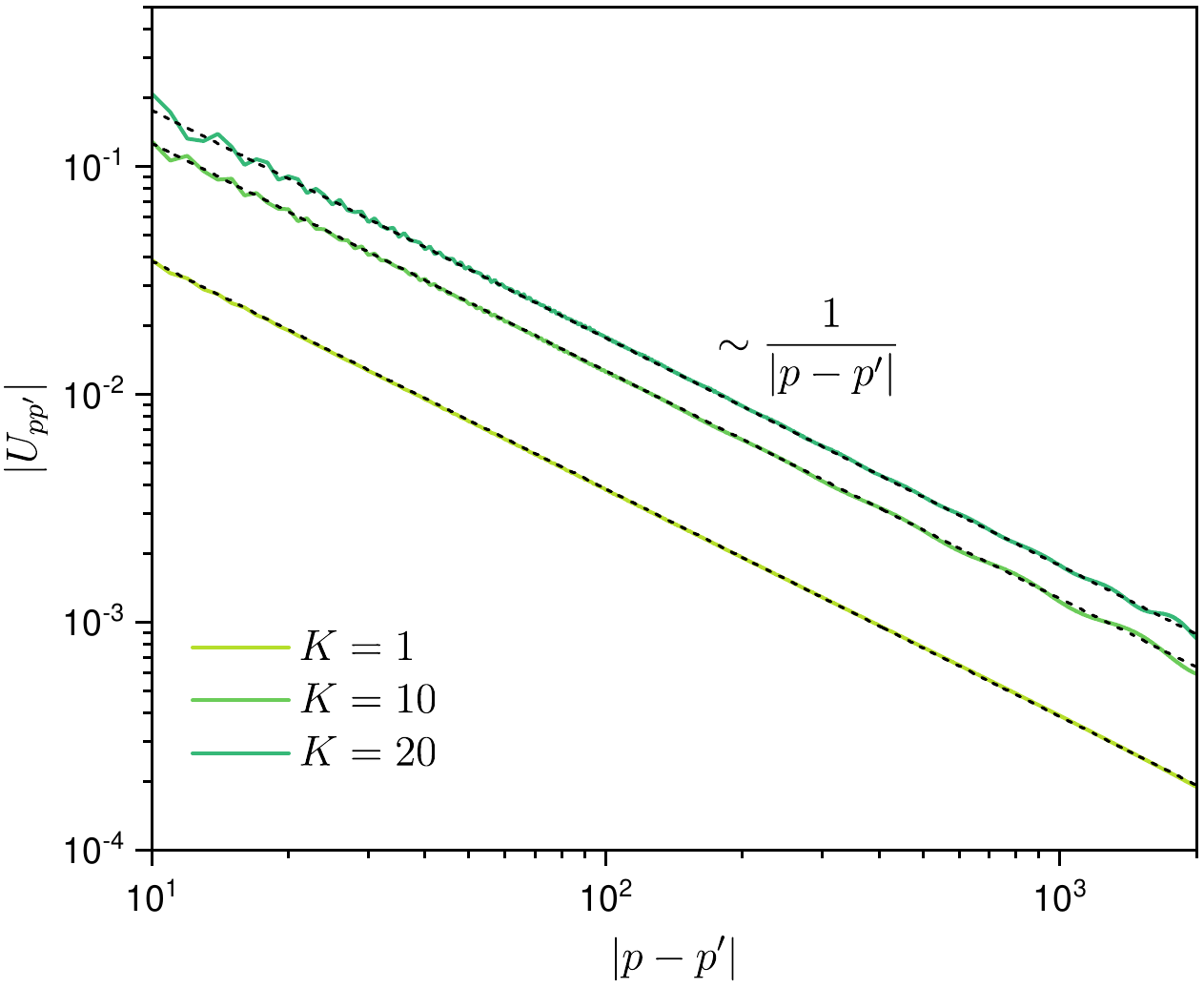}
\caption{\label{figUpp}The decay of off-diagonal matrix elements $|U_{pp^{\prime}}|$,  the dashlines are fitting with Eq. (\ref{eqA8}), verifying the conjecture even for large $K$. }
\end{figure}

\begin{figure}
\includegraphics[width=0.4\textwidth]{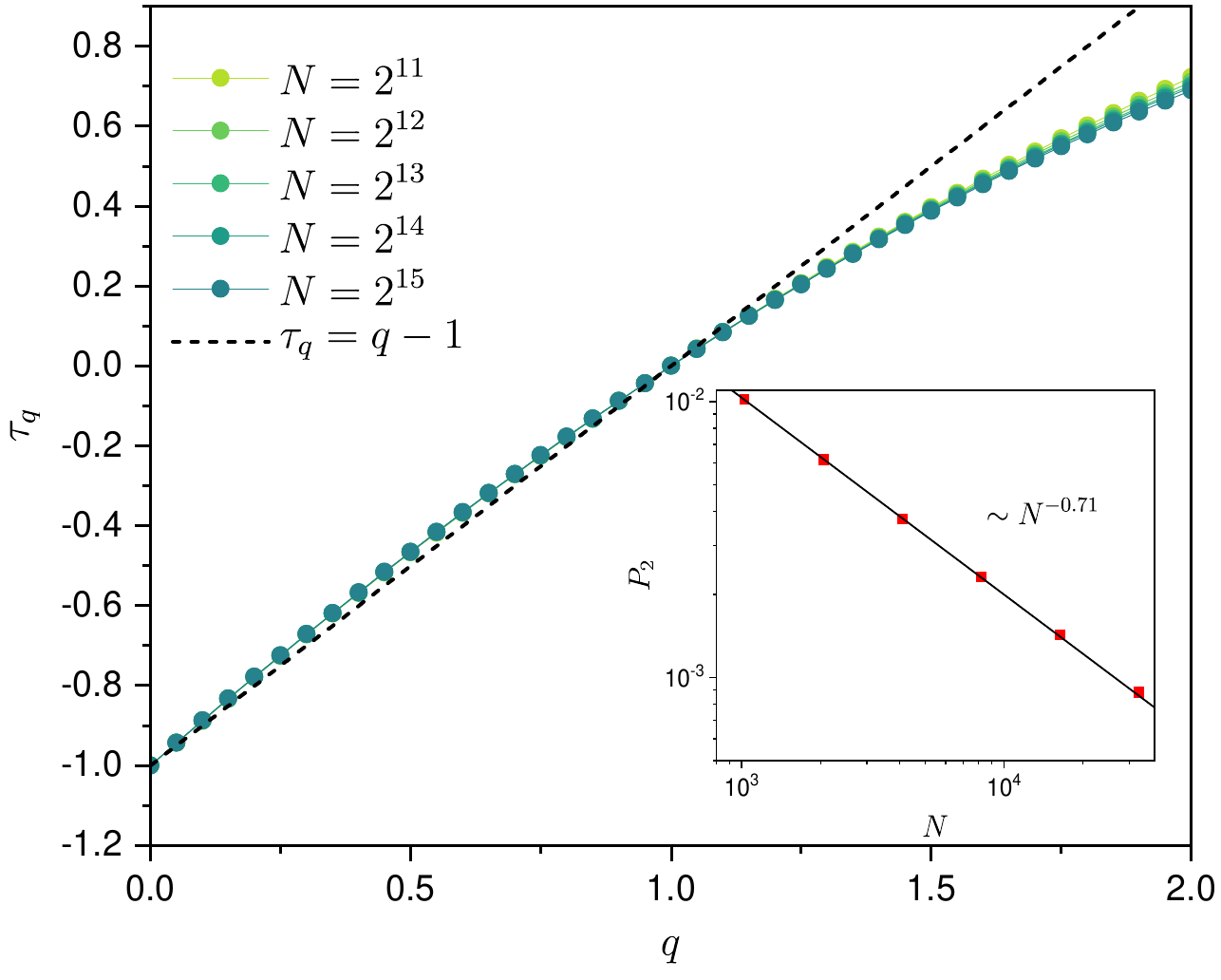}
\caption{\label{figTauq}Multifractal spectrum $\tau_q$ of the MKR model for $K=10$, showing clear deviations from the ergodic behavior. $\tau_q$ is determined through Eq.~\eqref{eq:tauq} for different system sizes, from $N=2^{11}$ to $N^{15}$. The data all lie on a single curve, which confirms the algebraic behavior of the eigenstate moments $\langle P_q \rangle$ with $N$. Inset: Inverse participation ratio $\langle P_2 \rangle \sim N^{-D_2}$ with $D_2 \approx 0.71$.}
\end{figure}

\begin{figure*}
         \includegraphics[width=0.4\linewidth]{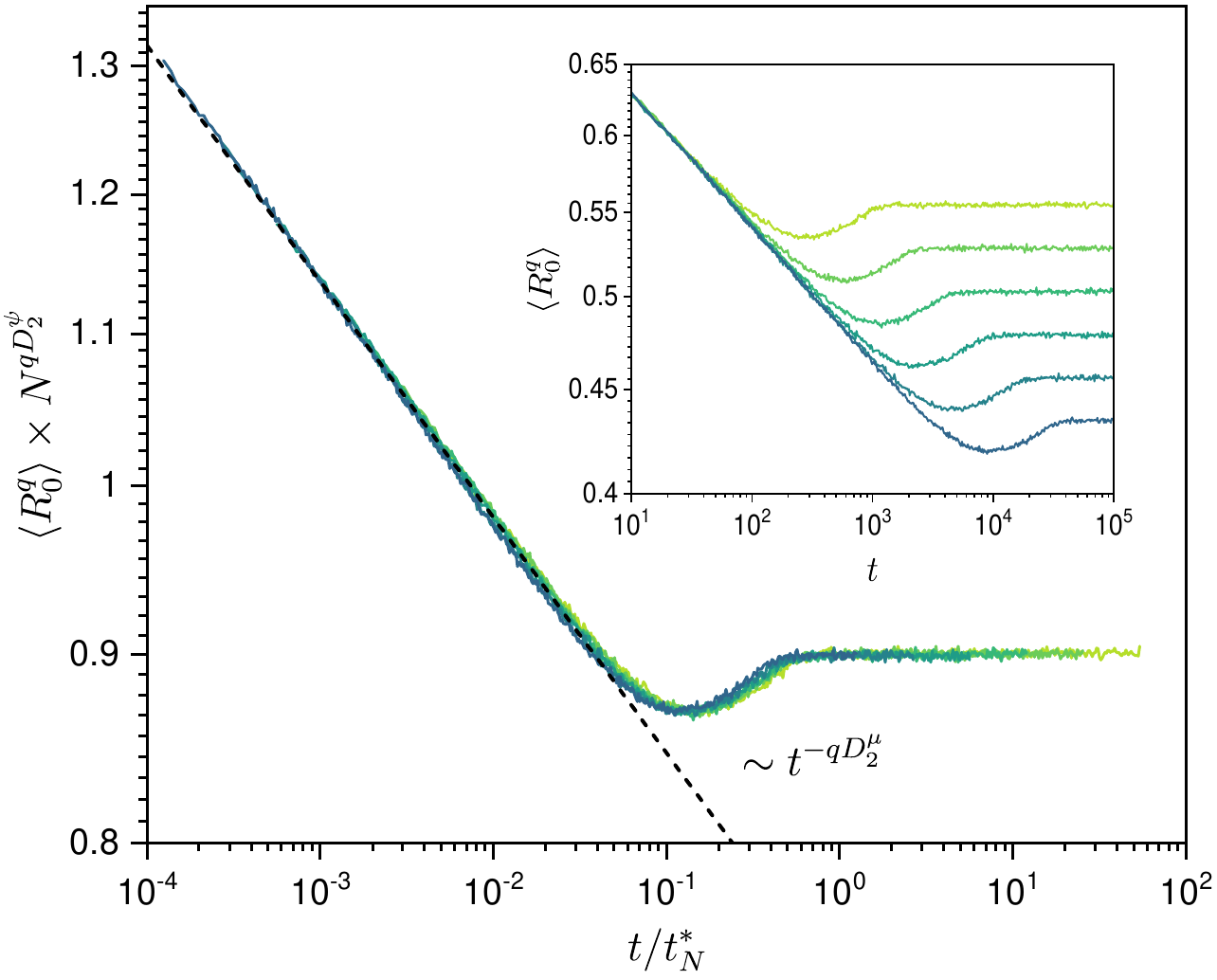}
         \includegraphics[width=0.4\linewidth]{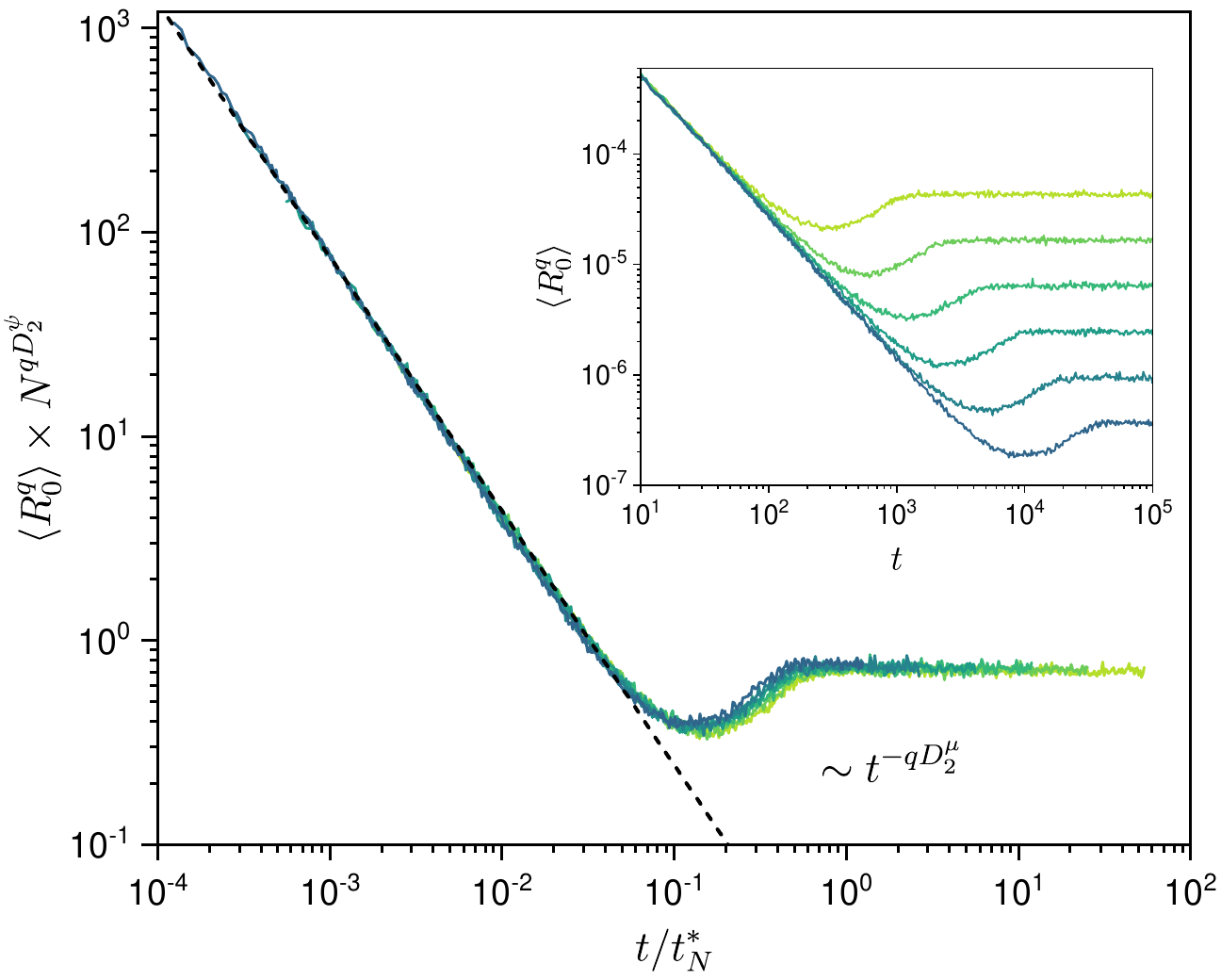}
    \caption{\label{figScaled_R0q_t}  Scaled  $\langle R_0^{q}\rangle \times N^{D_2^{q\psi}}$ as a function of the scaled time $t/t^*_N$ for two different $q$ values, $q=0.1$ for the upper panel and $q=2$ for the lower one, with the kicked strength $K=10$. The dashed line represents a fit of the dynamics by $\langle R_0^q\rangle\sim t^{-qD_2^{\mu}}$. Insets: $\langle R_0^q\rangle $  as a function of time $t$. Curves from the top to bottom correspond to system sizes $N=2^{10},2^{11},...,2^{15}$, respectively.}
\end{figure*} 

In this Appendix, we relate the power-law decay of the amplitudes of Floquet matrix elements $|U_{p, p^{\prime}}|$ in  momentum space with the singularity of the kicked potential $V(q)$ in real space. 

We consider the regime $K\ll1$ and make a first-order expansion in $K$ of the Floquet operator Eq.~\eqref{eq:Uop} as
\begin{equation}
\begin{split}
      U_{pp^{\prime}}=e^{i\Phi_{p}}\delta_{pp^{\prime}}-iKe^{i\Phi_{p}}\sum_{Q=1}^{N}F_{pQ}V(2\pi Q/N)F_{Qp^{\prime}}^{-1}.
\end{split}
\end{equation}
Therefore,
\begin{equation}
\begin{split}
      |U_{pp^{\prime}}|\simeq K|\sum_{Q=1}^{N}F_{pQ}V(2\pi Q/N)F_{Qp^{\prime}}^{-1}|
\end{split}
\end{equation}
for $p\neq p^{\prime}$.
Next, we evaluate the Fourier transform
\begin{equation}
\begin{split}
\sum_{Q=1}^{N}F_{pQ}V(2\pi Q/N)F_{Qp^{\prime}}^{-1}
&=\sum_{Q=1}^{N}\frac{1}{N}e^{2i\pi (p-p^{\prime})Q/N}V(\frac{2\pi Q}{N})
\end{split}
\end{equation}
as an integral. Notice that the potential $V(q)$ is symmetric with respect to $q=\pi$, hence,
\begin{equation}
\begin{split}
\frac{1}{2\pi}&\int_0^{2\pi}V(q)e^{i(p-p^{\prime})q}dq=\frac{1}{\pi}\int_0^{\pi}\ln(q)e^{i(p-p^{\prime})q}dq\\
&\underrel{|p-p^{\prime}|\rightarrow+\infty}{\sim}\frac{i}{|p-p^{\prime}|}\sum_{r=0}^{\infty}c_r(\ln|p-p^{\prime}|)^{1-r},
\end{split}    
\end{equation}
where the coefficients $c_r$ follow \cite{WONG1978173}:
\begin{equation}
\begin{split}
    c_r=(-1)^r\binom{1}{r}\sum_{k=0}^{r}\binom{r}{k}(\frac{\pi i}{2})^{(r-k)}.
\end{split}
\end{equation}
The dominating term is therefore:
\begin{equation}\label{eqA8}
    |U_{pp^{\prime}}|\sim\frac{1}{|p-p^{\prime}|}.
\end{equation}
Note that $V(q)$ has another singularity at  $q=\pi$, which is of higher order (first-derivative), compared to the singularity when $q=0( 2\pi)$ (zero order), therefore, we only take into account the lowest order. 

Although the above arguments are based on a first-order expansion in $K$ valid for $K\ll1$, numerical data presented in Fig.~\ref{figUpp} show that Eq. (\ref{eqA8}) is valid even for larger values of $K$.

\section{Multifractal properties of the MKR model}\label{appendixB}

Quantum multifractality can be characterized by the moments $P_q$, Eq.~\eqref{eq:Pq}, of eigenstate amplitudes $|\Psi_\alpha(i)|^2$, $\langle P_q\rangle\sim N^{-\tau_q}$,
where $\tau_q=D_q(q-1)$, $D_q$ are the multifractal dimensions and $N$ is the system size. Numerically, we compute $\tau_q$ by 
\begin{equation}\label{eq:tauq}
    \tau_q(N)=-[\log_2 \langle P_q(N)\rangle-\log_2 \langle P_q(N/2)\rangle].
\end{equation}
The numerical data are shown in Fig. (\ref{figTauq}) which confirm the multifractal properties of the MKR model, and in particular $D_2\approx 0.71$

\section{Two-parameter scaling properties of $\langle R_0^q\rangle$ and $\langle p^k\rangle$}\label{appendixC}

Fig.~(\ref{figScaled_R0q_t})  presents numerical data for $\langle R_0^q\rangle$ of the MKR model, Eq.~\eqref{eq2}, for different $q$ values, confirming the validity of the proposed two parameter scaling law Eq.~\eqref{eq:scaR0}.

\begin{figure*}[]
         \includegraphics[width=0.4\linewidth]{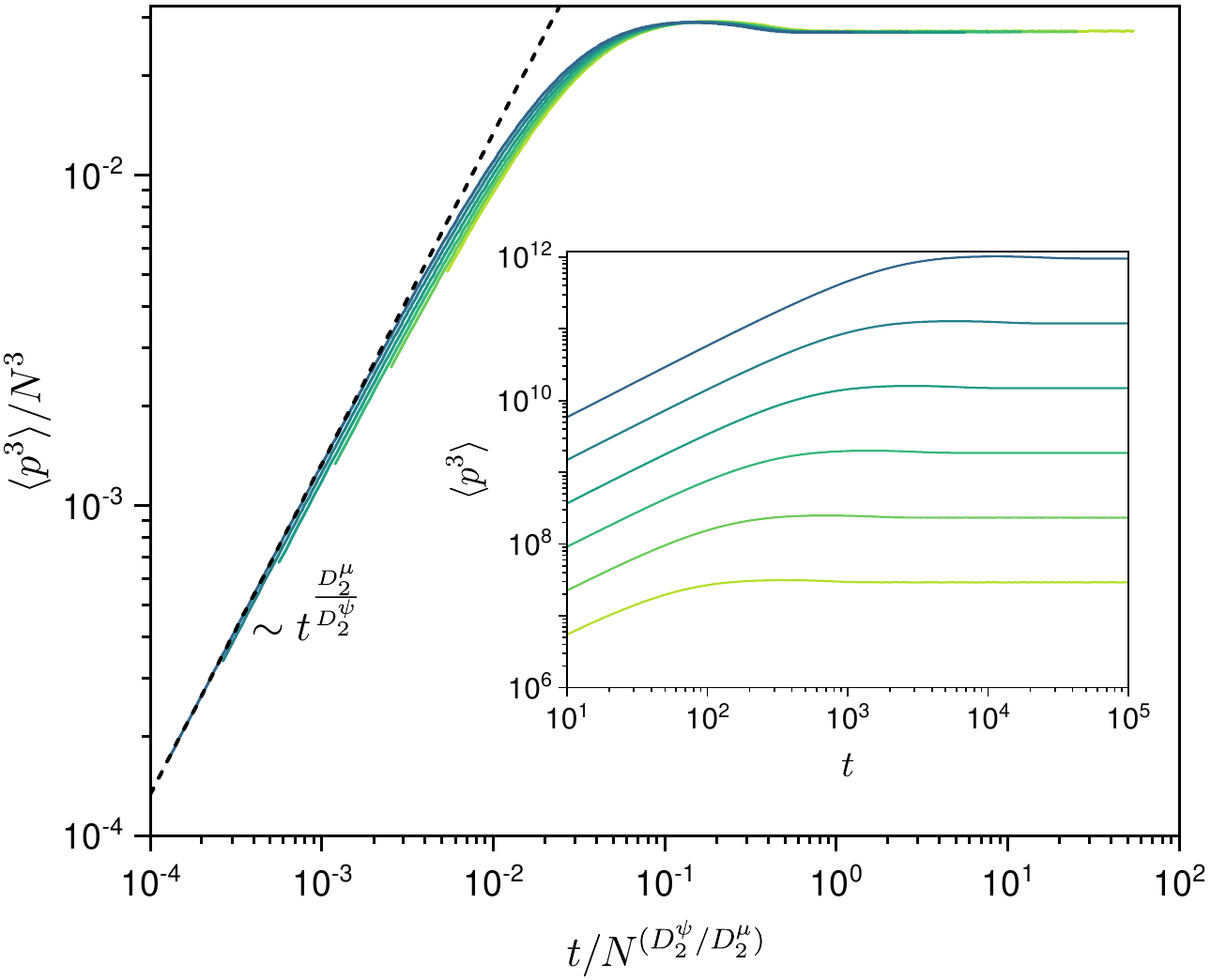}
         \includegraphics[width=0.4\linewidth]{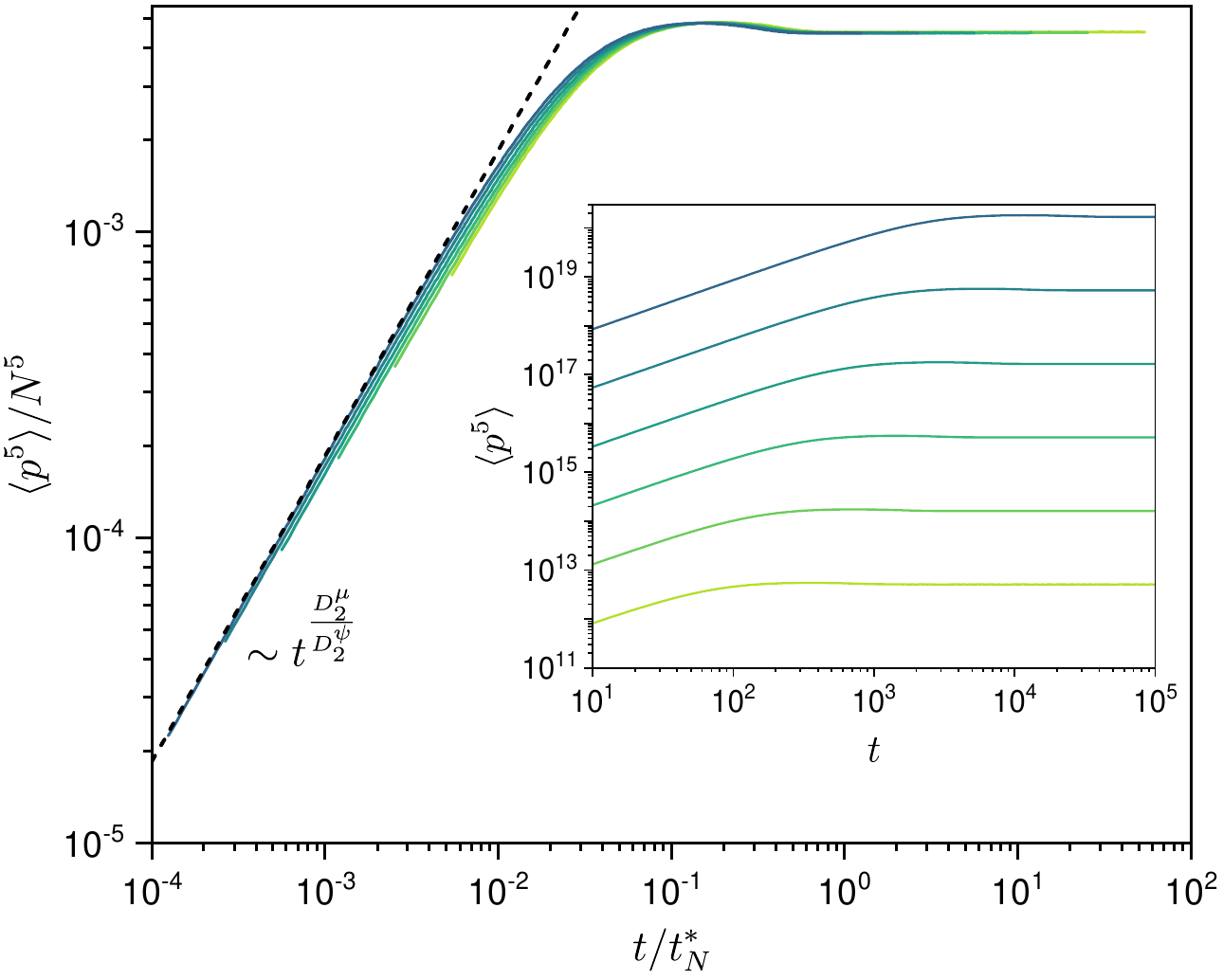}
    \caption{\label{figScaled_xk_t}  Scaled moments of the wave packet  $\langle p^k\rangle/N^k$ as a function of the scaled time $t/t^*_N$, for two different values of $k$, $k=3$ in the left panel and $k=5$ in the right panel. The dashed line represents a fit of the dynamics with $\langle p^{k}\rangle\sim t^{D_2^{\mu}/D_2^{\psi}}$. Inset: Moments of the wave packet  $\langle p^k\rangle$ as a function of the evolution time $t$. Curves from bottom to top correspond to system sizes $N=2^{10},2^{11},...,2^{15}$, respectively.}
\end{figure*} 

\begin{figure*}[]
         \includegraphics[width=0.4\textwidth]{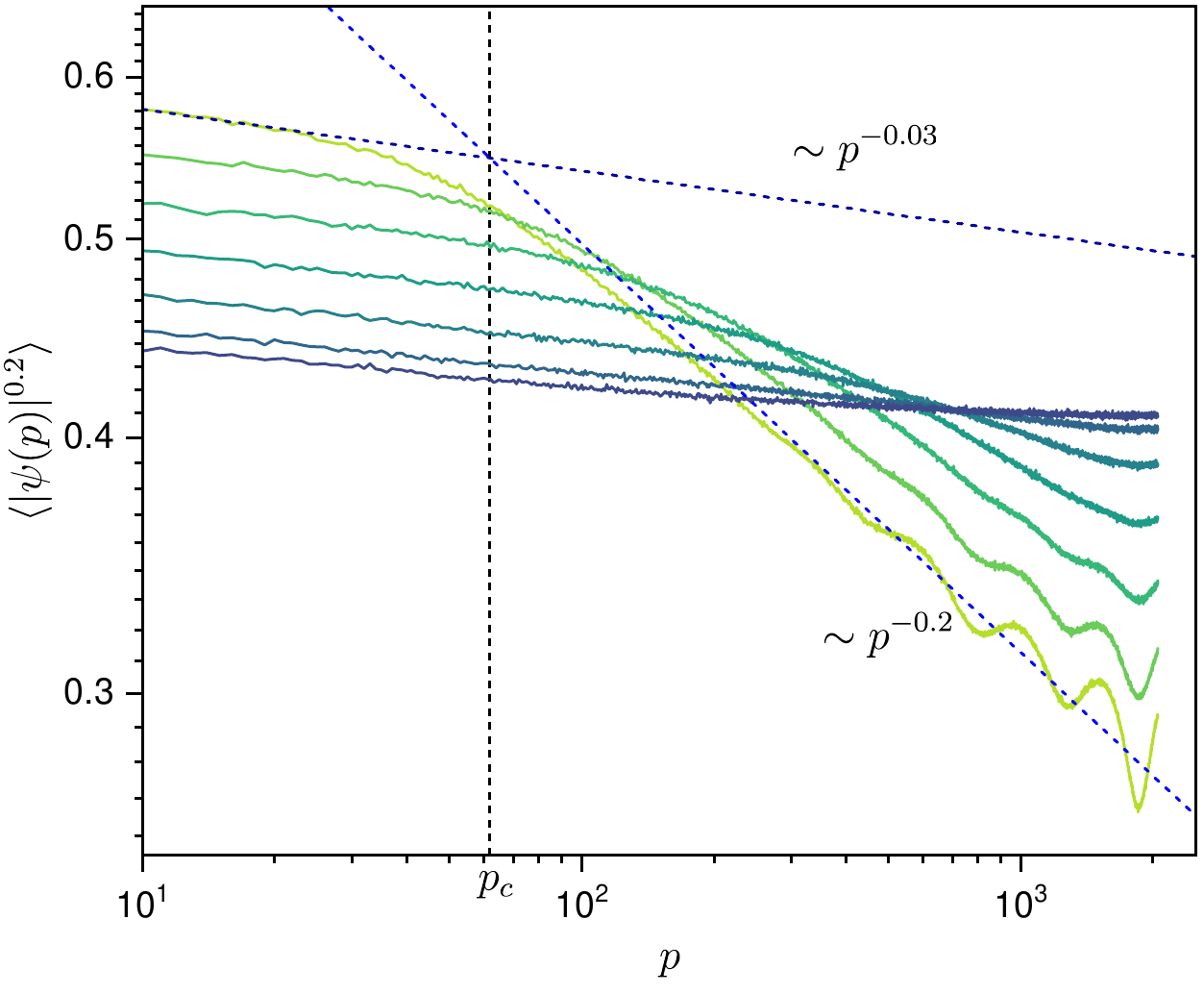}
         \includegraphics[width=0.4\textwidth]{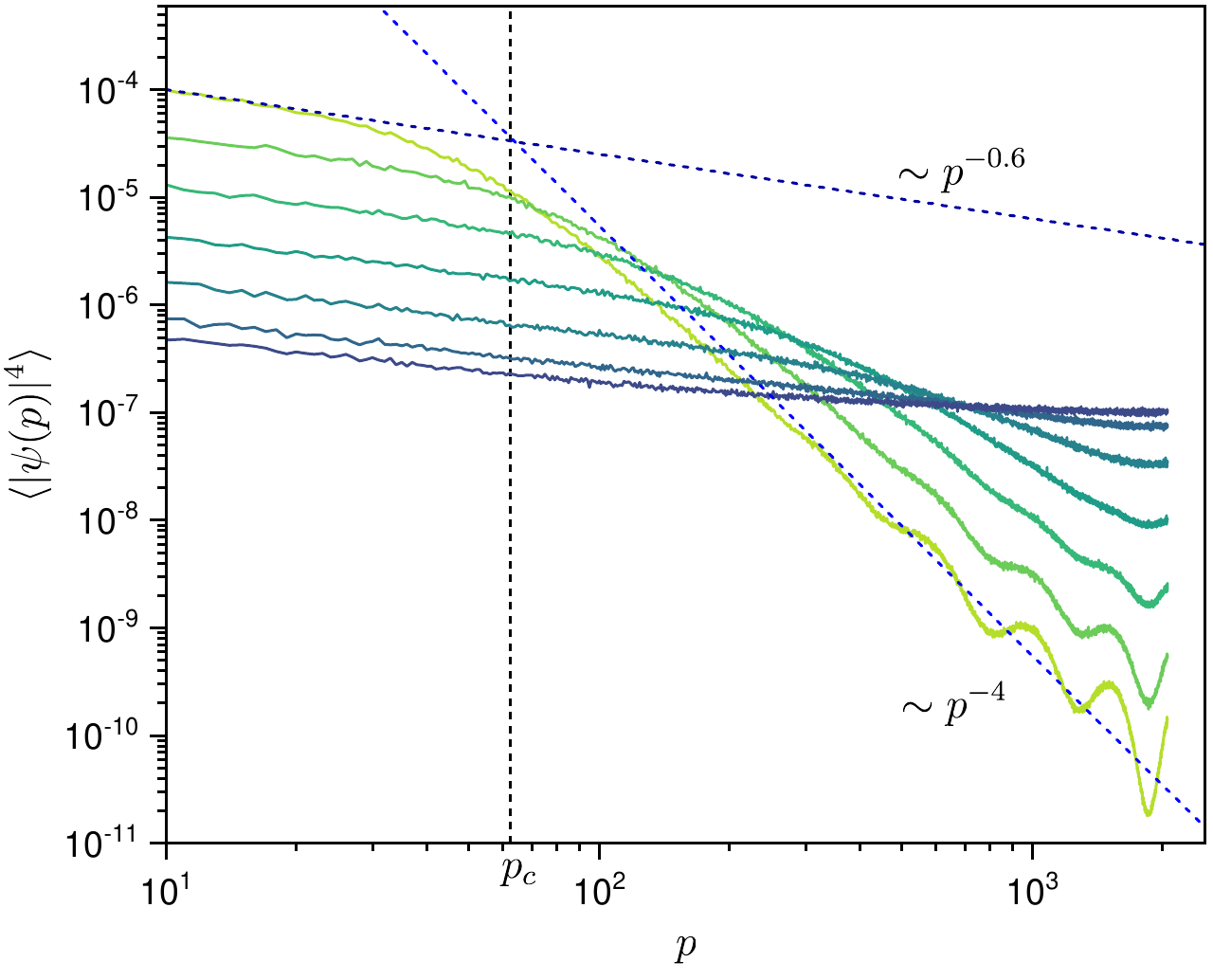}
    \caption{\label{figPsi2q} Averaged probability distribution $\langle|\psi(p,t)|^{2q}\rangle$ at different times for the MKR model with three different values of $q$, $q=0.2$ in the left panel and $q=4$ in the right panel. The dashed lines show the power-law behaviors corresponding to Eq.~(\ref{eq9}). Curves from top to bottom correspond to increasing time $t$. Here the kicked strength $K=10$ and the system size $N=2^{12}$.}
\end{figure*} 

\begin{figure*}[]
         \includegraphics[width=0.4\textwidth]{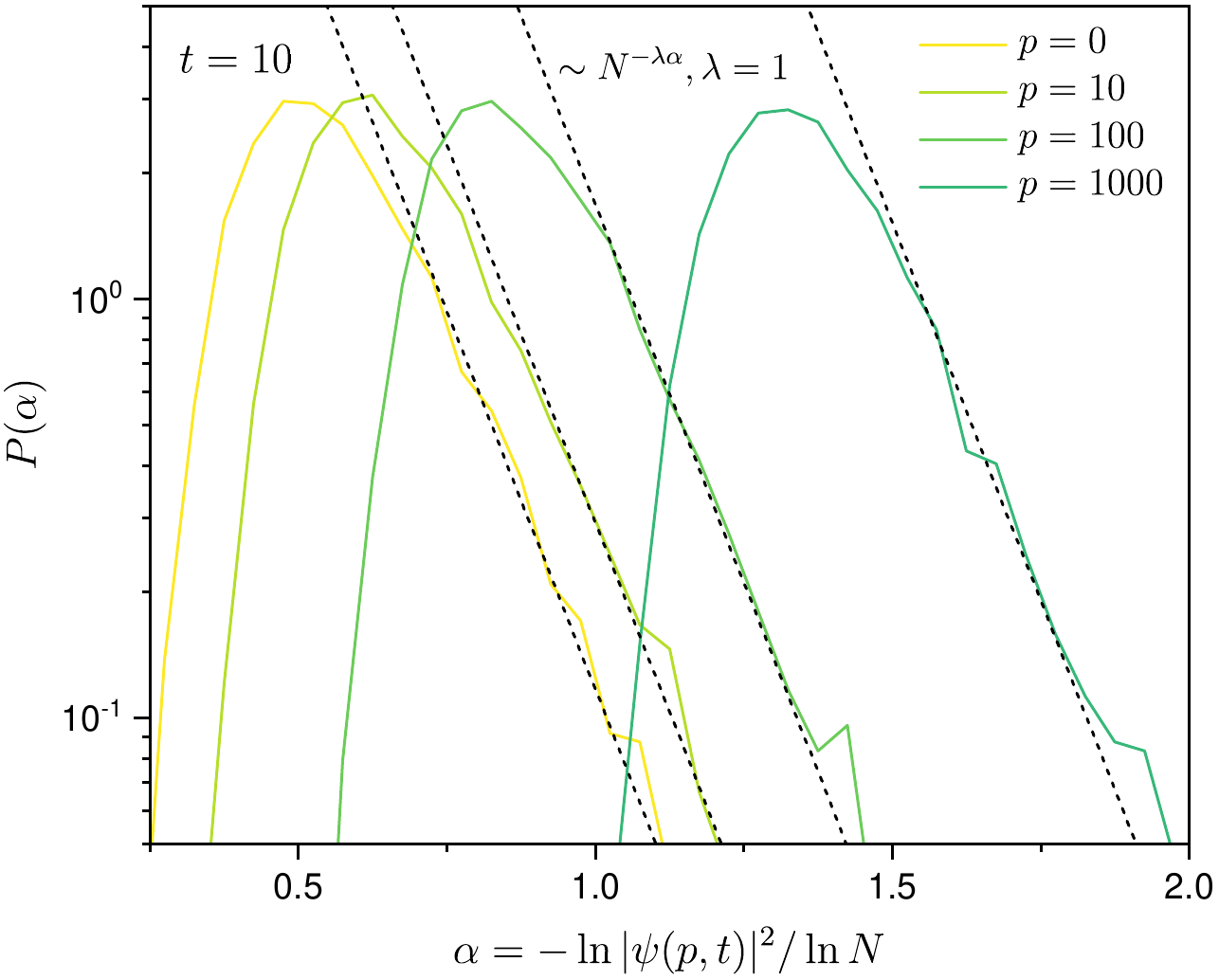}
         \includegraphics[width=0.4\textwidth]{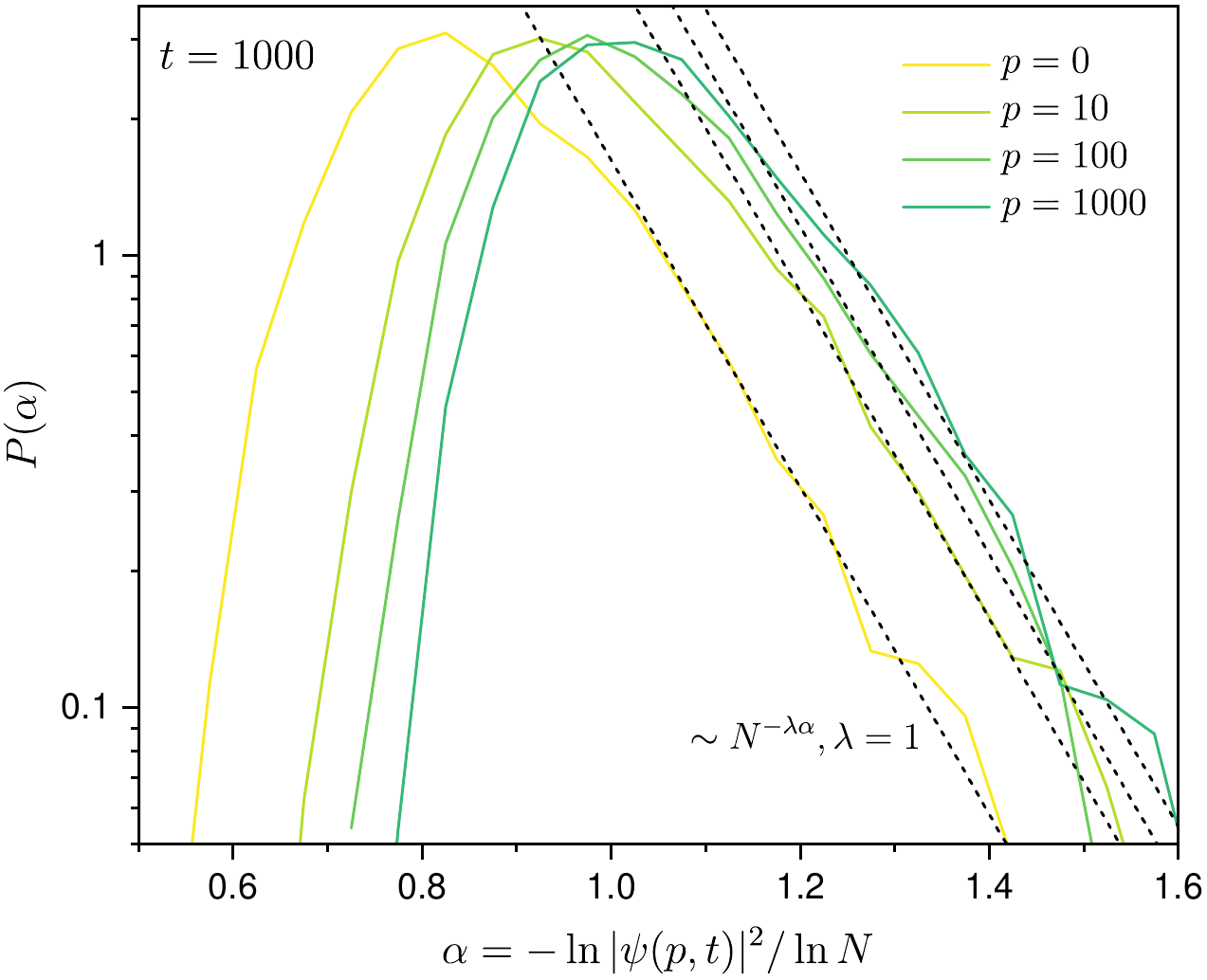}
    \caption{\label{figP_lnPsi2}Probability distributions of  $\alpha=-\ln|\psi(p,t)|^2/\ln N$ for the MKR model at various $p$ values from $p=0$ to $p=1000$, from right to left. In the left panel, we show data at time $t=10$, while $t=1000$ in the right panel. The system size is $N=2^{12}$. The dashed lines show fits of the exponential tails $P(\alpha)\sim N^{-\lambda\alpha}$ induced by Gaussian fluctuations and responsible for algebraic fat tails of the distribution of wave function amplitudes $\vert \psi\vert^2$ at small amplitudes. However, such an exponential tail is absent for small $\alpha$, i.e.~large wave function amplitudes $\vert \psi\vert^2$, the regime of interest when calculating positive moments $\langle \vert \psi(p,t)\vert\vert^{2q}\rangle$ with $q>0$.}
\end{figure*}

In Fig.~\ref{figScaled_xk_t}, we show numerical data for $\langle p^k\rangle$ for two different $k>1$ values, showing the universality of the prediction  $\langle p^{k}\rangle\sim t^{D_2^{\mu}/D_2^{\psi}}$ and the validity of the proposed scaling laws, Eq.~\eqref{eq17}.

\section{Average generalized wave packet $\langle|\psi(p,t)|^{2q}\rangle$ for different $q$ values and probability distribution of wave function amplitudes $\vert \psi \vert^2$}\label{appendixD}

\begin{figure*}[]
         \includegraphics[width=0.4\textwidth]{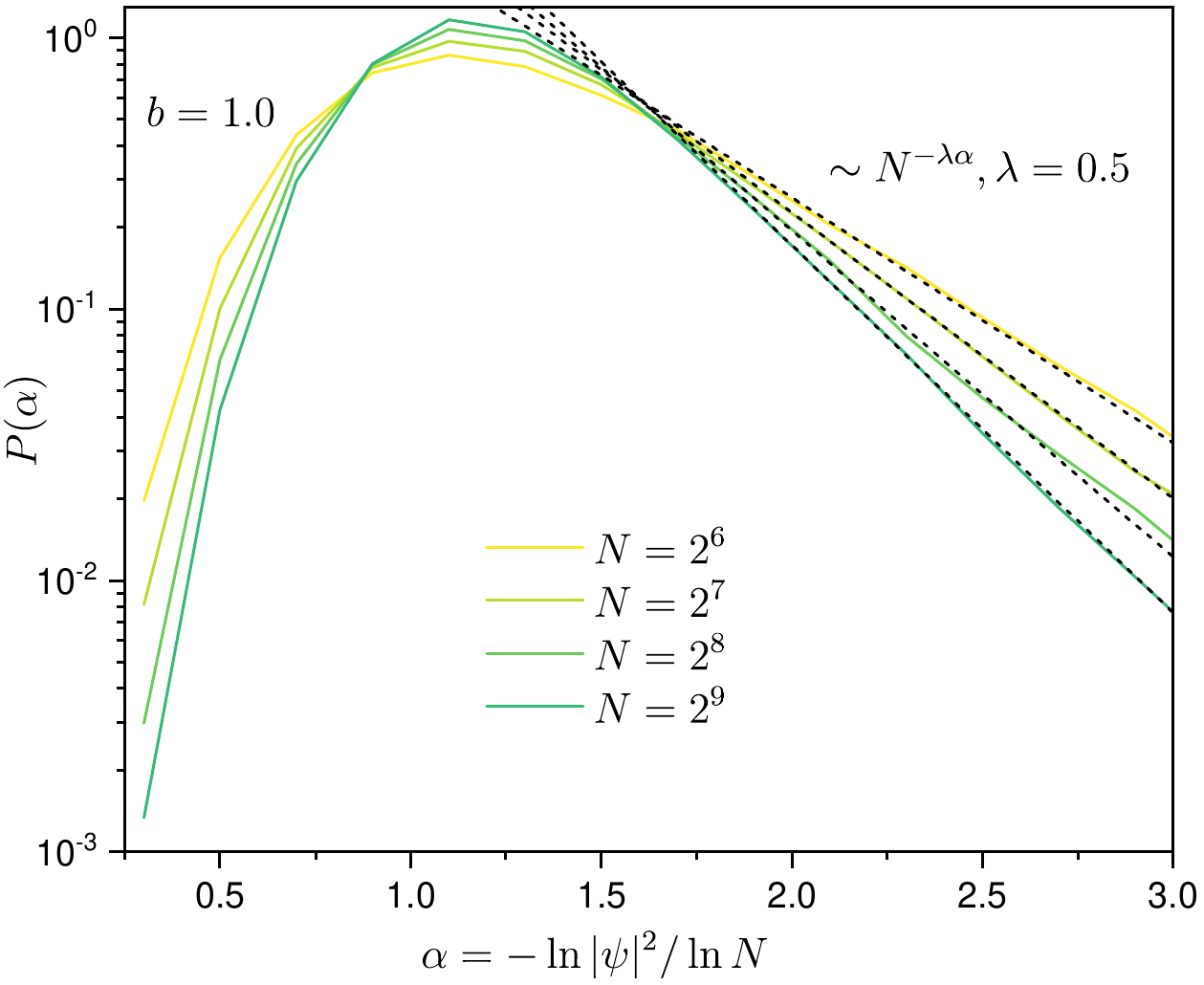}
    \caption{\label{figP_lnPsi2_PRBM}Probability distributions of  $\alpha=-\ln|\psi(p,t)|^2/\ln N$  for the critical PRBM model \cite{PhysRevE.54.3221, PhysRevB.62.7920}. The dashed lines represent fits of the exponential tails $P(\alpha)\sim N^{-\lambda\alpha}$ induced by Gaussian fluctuations. }
\end{figure*} 

Fig.~\ref{figPsi2q} presents numerical data for the generalized wave packets 
 $\langle|\psi(p,t)|^{2q}\rangle$, showing the same shape across different $q$ values, in particular the same multifractal wave-front $p_c$. 
 
 In Fig.~\ref{figP_lnPsi2}, we present the probability distribution of $\alpha=-\ln|\psi(p,t)|^2/\ln N$ for different $p$ and $t$ values. On the right side of the distribution corresponding to small wave function amplitudes $|\psi|^2$, we observe that there is an anomalously wide distribution $P(\alpha)\sim N^{-\lambda\alpha}$. Such distributions can be related to the Porter-Thomas law $P(\alpha)\sim N^{\beta(1-\alpha)/2}\exp(-\frac{1}{2}\beta N^{1-\alpha})$ as small amplitudes $|\psi(p,t)|^2$ are described by Random Matrix Theory \cite{PhysRev.104.483, PhysRevResearch.3.L022023}, where $\beta=1,2$ is the Dyson index corresponding to Orthogonal Ensemble and Unitary Ensemble. Hence, when $\alpha\gg 1 $ and $N\gg1$, $P(\alpha)\sim N^{-\lambda\alpha}$ with $\lambda=\beta/2$. We confirm such scaling behavior of Gaussian fluctuations both in the MKR model ($\beta=2$) and the critical PRBM model ($\beta=1$) \cite{PhysRevE.54.3221, PhysRevB.62.7920}, see Fig.~\ref{figP_lnPsi2_PRBM}.
 
 However, on the left side of the distribution corresponding to large wave function amplitudes $|\psi|^2$, $P(\alpha)$ decreases faster than exponentially, indicating the absence of an algebraic fat for the corresponding distribution of $|\psi(p,t)|^2$ at large amplitudes $|\psi(p,t)|^2$. This absence of large fluctuations at large amplitudes is responsible for $\langle|\psi(p,t)|^{2q}\rangle \sim \langle|\psi(p,t)|^2\rangle^q$ for $q>0$. Hence, the shape of $\langle|\psi(p,t)^{2q}|\rangle$ as a function of $p$ is the same for different $q>0$ values, in particular they have the same $p_c$.

\newpage
\bibliography{apssamp}

\end{document}